\documentclass[12pt]{article}
\usepackage{amsthm}
\usepackage{eurosym}
\usepackage{amsfonts}
\usepackage{amssymb,amsmath}
\usepackage[dvips]{color}
\usepackage{amscd}
\usepackage{tikz}
\usepackage{mathtools}
\usepackage{graphicx}
\usepackage{caption}
\usepackage{subcaption}

\newcommand{\xdownarrow}[1]{ {\left\downarrow\vbox to #1{}\right.\kern-\nulldelimiterspace} }
\usetikzlibrary{decorations.pathreplacing}

\usetikzlibrary{patterns}

\def\and{\mathrm{and}}

\newcommand{\ee}{\end{equation}}
\newcommand{\bea}{\begin{eqnarray}}
\newcommand{\eea}{\end{eqnarray}}
\newcommand{\beas}{\begin{eqnarray*}}
\newcommand{\eeas}{\end{eqnarray*}}
\newcommand{\ba}{\begin{array}}
\newcommand{\ea}{\end{array}}

\newcommand{\nbox}{{\,\lower0.9pt\vbox{\hrule \hbox{\vrule height 0.2 cm \hskip 0.19 cm \vrule height 0.2 cm}\hrule}\,}}

\def\href#1#2{#2}
\theoremstyle{plain}

\input{tcilatex}
\begin{document}


\newpage


\begin{titlepage}
\hfill
\vbox{
    \halign{#\hfil         \cr
           } 
      }  

\hbox to \hsize{{}\hss \vtop{ \hbox{}

}}

%

\vspace*{20mm}

\begin{center}

{\large \textbf{Interior analysis, stretched technique and bubbling geometries \vspace{0.4cm}}}

{\normalsize \vspace{10mm} }

{\normalsize {Qiuye Jia${}^{1}$}, Hai Lin${}^{2,3}$}

{\normalsize \vspace{10mm} }

{\small \emph{\textit{Mathematical Sciences Institute, Australian National University, Canberra ACT 2601, Australia
}} }

{\normalsize \vspace{0.2cm} }

{\small \emph{\textit{Shing-Tung Yau Center and School of Mathematics, Southeast University,\\
Nanjing 210096, China
}} }

{\normalsize \vspace{0.2cm} }

{\small \emph{\textit{Yau Mathematical Sciences Center, Tsinghua University,
Beijing 100084, China
\\
}} }
{\normalsize \vspace{0.4cm} }
\end{center}


\begin{abstract}

{\normalsize \vspace{0.3cm} }

We perform a detailed analysis of quarter BPS bubbling geometries with AdS asymptotics and their
corresponding duality relations with their dual states in the quantum field theory side, among other aspects. We derive generalized Laplace-type equations with sources, obtained from linearized Monge-Ampere equations, and used for asymptotically AdS geometry. This enables us
to obtain solutions specific to the asymptotically AdS context. We conduct a thorough analysis of
boundary conditions and explore the stretched technique where boundary conditions are imposed on a
stretched surface. These boundary conditions include grey droplets. This stretched technique is naturally used for the superstar, where we place grey droplet boundary conditions on the stretched surface. We also perform a coarse-graining of configurations and analyze the symplectic forms on the configuration space and their coarse-graining.

\end{abstract}

\end{titlepage}

\vskip 1cm \newpage \newpage \newpage \newpage

\section{Introduction}

\label{sec_introduction}

The gauge-gravity correspondence has revealed remarkable relations and
equivalences between ordinary quantum systems on one side and quantum
gravity on the other side, e.g. \cite%
{Beisert:2010jr,Rangamani:2016dms,Horowitz:2006ct} and other related
aspects. In the correspondence, the spacetime geometries are emergent from
the dynamics of ordinary quantum systems on the dual side.

On the gravity side, there are giant graviton branes \cite{McGreevy:2000cw}-%
\cite{Lin:2004nb}. The corresponding states in the Hilbert space of the
quantum field theory are explicitly mapped to the states in the gravity
side. The dual large operators as well as their operator bases have been
illuminated \cite{Corley:2001zk}-\cite{Lin:2004nb}. These are brane-like
operators, which are holographically dual to branes on the gravity side.
Moreover, these branes are associated to bubbling geometries, in the bulk
side. Both giant graviton branes and bubbling geometries have string
excitations. Analyses in the field theory side show that these different
states live in the same Hilbert space. See also related interesting
discussions, e.g. \cite{Gaiotto:2021xce}-\cite{Beccaria:2023hip}.

These smooth bubbling geometries correspond to excited states in the Hilbert
space of the dual quantum field theory. The states in the Hilbert space of
the quantum field theory are explicitly mapped to the states in the gravity
side. The Hilbert spaces of states in both sides are isomorphic to each
other. The configuration spaces of the droplet configurations have been
analyzed and illuminated in, e.g. \cite%
{Berenstein:2017abm,Balasubramanian:2005mg,Grant:2005qc,Maoz:2005nk}.
Various types of degrees of freedom on the gravity side or on the string
theory side, can be described by corresponding operators on the quantum
field theory side. Moreover, different operator bases can be transformed
into each other by a change of basis.

There are various operator bases for large operators, corresponding to heavy
states in the gravity. The large operators includes those describe giant
gravitons and emergent geometries, among other things. The large operators
also describe further excitations on these heavy excited states. The
correlation functions between light operators and large operators can also
be computed. The large operators can be expanded systematically in terms of
representation bases. For more information along these ideas and aspects,
see e.g. \cite{Bhattacharyya:2008rb}-\cite{Lewis-Brown:2020nmg}. Various
bases can be transformed into each other. These heavy operators correspond
to heavy excited states in the Hilbert space of the quantum gravity side.

One can also compute the interactions of these objects in the gravity side,
by computing correlation functions of their dual operators. For example, the
correlation functions of giant gravitons and light fields have been computed
from gravity and from gauge theory sides, and they match with each, e.g.
\cite{Bissi:2011dc}-\cite{Holguin:2022zii}. The operators used in the
computations are Schur operators and their superpositions. It has been very
useful computing correlation functions involving large operators, e.g. \cite%
{Abajian:2023jye,Holguin:2023orq,Holguin:2022zii}, for strong gravitation in
AdS, among other things. Transformations between bases are useful for
computations, including those aspects related to amplitudes of the gravity
side. These heavy operators involve emergent geometries in the dual quantum
gravity system. These excited states of geometries live in the same quantum
Hilbert space. One may further deform the Hamiltonians while keeping
structures in the Hilbert space. The set-up on the field theory side is very
calculable, and this enables us to go into further and more sophisticated
regimes of the duality.

Other examples or classes of bubbling geometries with quarter BPS and eighth
BPS cases have been analyzed in e.g. \cite{Chen:2007du}-\cite{Lunin:2007ab},
e.g. for the $N$=4 SYM example, among other things. There are other
less-supersymmetric deformations, e.g. \cite{Liu:2004ru} and related
discussions. One can obtain the symplectic form for the configuration of the
quarter BPS geometries, by using the same procedure as in \cite%
{Grant:2005qc,Maoz:2005nk,Balasubramanian:2018yjq}. Moreover, by discrete
identification of base space, one can obtain orientifold cases, e.g. \cite%
{Mukhi:2005cv}-\cite{Caputa:2013vla} and their related aspects. Various
other aspects have also been discussed and addressed, e.g. \cite%
{Bhattacharyya:2010yg}-\cite{Balasubramanian:2007qv}. There are many other
related analysis and aspects. See also discussions on related giant
gravitons, e.g. \cite{Biswas:2006tj,Mandal:2006tk,Kim:2006he} and their
closely related aspects.

The correlators with heavy operators have been considered from various
perspectives. It is particularly interesting to analyze the HHLL
correlators. One can also compute those HHLL correlators, where heavy H
operators are dual to background bubbling geometry and light L operators are
dual to light fields in AdS. Fluctuations of light fields on these
backgrounds can also be captured by those types of very interesting
correlators. Very interesting and similar types of correlators involving
heavy and light operators have been computed recently in e.g. \cite%
{Paul:2023rka,Caetano:2023zwe,Li:2019tpf}. HHLL correlators can also capture
the string degrees of freedom in bubbling geometries, e.g. \cite%
{deMelloKoch:2009jc,Berenstein:2020jen,Chen:2007gh,deMelloKoch:2018ert} and
related references.

Heavy geometries are also dual to Wilson operators and surface operators.
There are also bubbling geometries for line operators, e.g. \cite%
{Yamaguchi:2006te}-\cite{Gomis:2008qa} and for surface operators, e.g. \cite%
{Gukov:2008sn}-\cite{Drukker:2008wr}, as examples, and various their related
references. There are many interesting and closely related phenomena, among
these cases. In the context of Wilson operators, the involved path integral
can be computed by powerful localization formalism. The localization
techniques are also very useful for aspects of local operators, e.g. \cite%
{Berenstein:2022srd,Asano:2014eca,Asano:2012zt} and related references. It
would be nice to have an unified understanding of various interconnected
aspects.

Hence, here, we analyze in more details the quarter BPS bubbling geometries
and their properties and duality relations with their dual states on the
quantum field theory side, among many other aspects. In Sec. 2, we obtain
generalized Laplace-type equations around reference geometries. In Sec. 3,
we analyze solutions with AdS asymptotics. In Sec. 4, we analyze in details
solutions to the general equations with boundary conditions. In Sec. 5, we
analyze the stretched technique where we place boundary conditions on the
stretched surface. In Sec. 5.2 and 5.3, we make coarse graining of
configurations, with grey droplet boundary conditions, and analyze
coarse-graining for superstar geometries, among other things. Finally in
Sec. 6, we conclude and discuss closely related aspects. More details of
computations are included in Appendices A to F.

\section{Linearized equation with sources}

\label{sec 2} \renewcommand{\theequation}{2.\arabic{equation}} %
\setcounter{equation}{0} \renewcommand{\thethm}{2.\arabic{thm}} %
\setcounter{thm}{0} \renewcommand{\theprop}{2.\arabic{prop}} %
\setcounter{prop}{0}

\subsection{Linearized equation on reference geometries}

We consider smooth quarter BPS bubbling geometries as described in the
Introduction section. We derive a linearized equation in 5d (parametrized by
$y,z_{i},z_{\bar{j}},i,j=1,2$). Here, in the same way, there are multiple
droplets in quarter BPS bubbling geometries. The 10D metric is
\begin{eqnarray}
&&ds_{10}^{2}=-h^{-2}(dt+\omega )^{2}+h^{2}((Z+\frac{1}{2})^{-1}2h_{i\bar{j}%
}dz^{i}d\bar{z}^{\bar{j}}+dy^{2})+y(e^{G}d\Omega _{3}^{2}+e^{-G}(d\psi
+A)^{2}),  \notag  \label{GenericQuarterBPS} \\
&&h^{-2}=2y\cosh G,\qquad h_{i\bar{j}}=\partial _{i}\partial _{\bar{j}}K.
\end{eqnarray}%
We have $Z\equiv \frac{1}{2}\mathrm{\tanh }G=-\frac{1}{2}y\partial _{y}\frac{%
1}{y}\partial _{y}K$. The four-dimensional base metric $h_{i\bar{j}}$ is
Kahler, and satisfies a Monge-Ampere type equation, and an auxiliary
condition, see also \cite{Chen:2007du,Lunin:2008tf},
\begin{eqnarray}
&&\log \det h_{i\bar{j}}=\log (Z+\frac{1}{2})+n\eta \log y+\frac{1}{y}%
(2-n\eta )\partial _{y}K+D(z_{i},\bar{z}_{\bar{j}}),  \notag \\
&&(1+\ast _{4})\partial \bar{\partial}D=\frac{4}{y^{2}}(1-n\eta )\partial
\bar{\partial}K.  \label{eq:MA2}
\end{eqnarray}%
Here, the range of $y$ is $0\leq y<\infty .$

Consider the case $n\eta =1$, $D=\mathrm{const}$. The solutions for $%
AdS_{5}\times S^{5}$ are in e.g. (6.19) of \cite{Chen:2007du}. Other smooth
solutions are characterized by smooth droplet boundary conditions. The
equation, for $n\eta =1$, $D=\mathrm{const.}$, is%
\begin{eqnarray}
\det \partial _{i}\partial _{\bar{j}}K &=&(Z+\frac{1}{2})e^{\frac{1}{y}%
\partial _{y}K+\frac{1}{2}\log y^{2}+D},  \notag \\
Z &\equiv &-\frac{1}{2}y\partial _{y}\frac{1}{y}\partial _{y}K.
\label{eqn a}
\end{eqnarray}%
Here $D$ is a constant and can be related to an irrelevant integration
constant. $\frac{1}{2}\log y^{2}+D$ has no $z_{i},z_{\bar{j}}$ dependence
and can be associated to an integration constant for $K$. The RHS only
involves derivative with respect to $y^{2}$, and the LHS only involves
derivative with respect to $z_{i},z_{\bar{j}}$. There also exist scaling
transformations and shift transformations for the Kahler potential \cite%
{Kimura:2011df,Chen:2007du}.

At $y=0$, we have $Z-\frac{1}{2}=-1$ for black droplets where $S^{3}$
shrinks to zero smoothly. The $S^{3}$ combines with a radial coordinate to
form locally flat space when shrinking smoothly. At $y=0$, we have $Z-\frac{1%
}{2}=0$ for white droplets where $S^{1}$ shrinks to zero smoothly. The $%
S^{1} $ combines with a radial coordinate to form locally flat space when
shrinking smoothly. White droplets are the complementary regions of the
black droplets. The ten dimensional geometry is smooth \cite{Chen:2007du}.
There are quantized flux quanta through various interior cycles in the
smooth geometry, and the geometries have complicated topologies and are
everywhere smooth \cite{Chen:2007du,Kimura:2011df,Lunin:2008tf}.

There is certain complementarity between white droplets and black droplets
in the interior region. When the black droplets become smaller, the white
droplets in between them may become bigger and vice versa, since in the
interior droplet space, the white droplets are the complementary regions of
the black droplets. When you define one region, the other is its complement,
and vice versa. In fact, the white region can be defined as the complement
of the black region. We may redefine a function $u(z)_{y=0}=\frac{1}{2}%
-Z(z,y)_{y=0}$, and then $u(z)$ is 1 on black regions and 0 on white
regions. These smooth geometries correspond to pure states of the theory.

Consider the solution for $AdS_{5}\times S^{5}$ as $K_{(0)}$. We want to get
a special family of other solutions by adding sources $\delta _{i}\ $on the $%
y=0$ space (parametrized by $z_{i},z_{\bar{j}},i,j=1,2$), on top of the $%
AdS_{5}\times S^{5}$ background. This idea has not been proposed before. We
superimpose these sources for the function $Z$.

The solution for $AdS_{5}\times S^{5}$ is as follows. Here $%
r^{2}=|z_{1}|^{2}+|z_{2}|^{2}$. $K_{(0)}^{\prime }=\partial _{r^{2}}K_{(0)}={%
B/}2{r^{2}}$, which can be integrated to give the Kahler potential
\begin{equation}
K_{(0)}(r^{2},y)=\frac{1}{2}\int^{r^{2}}\frac{{B(r^{2},y)}}{{r^{2}}}d(r^{2}).
\end{equation}%
\begin{equation}
B=\frac{1}{2}(r^{2}-y^{2}-1)+\sqrt{\frac{1}{4}(r^{2}+y^{2}-1)^{2}+y^{2}}.
\end{equation}

The Kahler potential is
\begin{eqnarray}
K_{(0)} &=&\frac{1}{2}\left( \frac{1}{2}(r^{2}+y^{2}+1)+\sqrt{\frac{1}{4}%
(r^{2}+y^{2}-1)^{2}+y^{2}}\right)  \notag \\
&&-\frac{1}{2}\log \left( \frac{1}{2}(r^{2}+y^{2}+1)+\sqrt{\frac{1}{4}%
(r^{2}+y^{2}-1)^{2}+y^{2}}\right)  \notag \\
&&-\frac{1}{2}y^{2}\log \left( \frac{1}{2}(-r^{2}+y^{2}+1)+\sqrt{\frac{1}{4}%
(r^{2}+y^{2}-1)^{2}+y^{2}}\right) +\frac{1}{2}y^{2}\log (y).  \notag
\label{eq:kpot} \\
&&
\end{eqnarray}%
For more details, see Appendix A.

Now consider a family of new solutions
\begin{equation}
K=K_{(0)}+sK_{(1)},  \label{eqn 05}
\end{equation}%
where $s$ is a small parameter. The $y=0$ is a four-dimensional droplet
space. Here, the $K_{(0)}$ is general and is not restricted to be the $AdS$
solution. We cover the source $\delta _{i}$ by a four-dimensional
ball-neighborhood $B_{i}$. We use an enclosure surface which is along the $y$
direction and ends on this four-dimensional ball-neighborhood. The above
solution is good for the region outside these enclosure surfaces; these
regions are where the change of the metric due to added sources are small;
The norm of $s\partial _{i}\partial _{\bar{j}}K_{(1)}$ is much smaller than
the norm of $\partial _{i}\partial _{\bar{j}}K_{(0)}$ in these regions. The
white region for $Z=\frac{1}{2},y=0$ on the droplet space is now $%
R^{4}\backslash \cup _{i}B_{i}$. The region $B_{i}$ are sources with $Z=-%
\frac{1}{2},y=0$. In other words, $B_{i}$ are `$Z=-\frac{1}{2}$' sources of
black regions. We make droplet configurations as the collections of these `$%
Z=-\frac{1}{2}$' sources. The $B_{i}$ are domains on the $y=0$ space.

Now we derive 5d linearized equation for $K_{(1)}$, from (\ref{eqn a}), for
the region outside the enclosure surfaces. We keep terms up to $o(s)$ terms.
The order $o(s^{0})$ terms cancel precisely, due to the reference geometry.
We have:
\begin{eqnarray}
\det \partial _{i}\partial _{\bar{j}}K &=&\partial _{1}\partial _{\bar{1}%
}K\partial _{2}\partial _{\bar{2}}K-\partial _{1}\partial _{\bar{2}%
}K\partial _{2}\partial _{\bar{1}}K  \notag \\
&=&\det \partial _{i}\partial _{\bar{j}}K_{(0)}  \notag \\
&&+s\left( \partial _{1}\partial _{\bar{1}}K_{(0)}\right) \partial
_{2}\partial _{\bar{2}}K_{(1)}+s\left( \partial _{2}\partial _{\bar{2}%
}K_{(0)}\right) \partial _{1}\partial _{\bar{1}}K_{(1)}  \notag \\
&&-s\left( \partial _{1}\partial _{\bar{2}}K_{(0)}\right) \partial
_{2}\partial _{\bar{1}}K_{(1)}-s\left( \partial _{2}\partial _{\bar{1}%
}K_{(0)}\right) \partial _{1}\partial _{\bar{2}}K_{(1)}
\end{eqnarray}%
and%
\begin{eqnarray}
&&\frac{1}{2}(1-4y^{2}\partial _{y^{2}}\partial _{y^{2}}K)e^{2\partial
_{y^{2}}K+\frac{1}{2}\log y^{2}+D}  \notag \\
&=&\frac{1}{2}(1-4y^{2}\partial _{y^{2}}\partial _{y^{2}}K_{(0)})e^{\frac{1}{%
2}\log y^{2}+D}e^{2\partial _{y^{2}}K_{(0)}}  \notag \\
&&+s[(1-4y^{2}\partial _{y^{2}}\partial _{y^{2}}K_{(0)})\partial
_{y^{2}}K_{(1)}-2y^{2}\partial _{y^{2}}\partial _{y^{2}}K_{(1)}]e^{2\partial
_{y^{2}}K_{(0)}+\frac{1}{2}\log y^{2}+D}.
\end{eqnarray}%
Hence (denoting $V=K_{(1)}$),
\begin{eqnarray}
&&\left( \partial _{2}\partial _{\bar{2}}K_{(0)}\right) \partial
_{1}\partial _{\bar{1}}V+\left( \partial _{1}\partial _{\bar{1}%
}K_{(0)}\right) \partial _{2}\partial _{\bar{2}}V  \notag  \label{K1 eqn 02}
\\
&&-\left( \partial _{1}\partial _{\bar{2}}K_{(0)}\right) \partial
_{2}\partial _{\bar{1}}V-\left( \partial _{2}\partial _{\bar{1}%
}K_{(0)}\right) \partial _{1}\partial _{\bar{2}}V  \notag \\
&=&e^{2\partial _{y^{2}}K_{(0)}+\frac{1}{2}\log y^{2}+D}[(1-4y^{2}\partial
_{y^{2}}\partial _{y^{2}}K_{(0)})\partial _{y^{2}}V-2y^{2}\partial
_{y^{2}}\partial _{y^{2}}V].  \label{K1 eqn}
\end{eqnarray}

This form of linearized equation can also be written as%
\begin{eqnarray}
&&\left( \partial _{2}\partial _{\bar{2}}K_{(0)}\right) \partial
_{1}\partial _{\bar{1}}V+\left( \partial _{1}\partial _{\bar{1}%
}K_{(0)}\right) \partial _{2}\partial _{\bar{2}}V-\left( \partial
_{1}\partial _{\bar{2}}K_{(0)}\right) \partial _{2}\partial _{\bar{1}%
}V-\left( \partial _{2}\partial _{\bar{1}}K_{(0)}\right) \partial
_{1}\partial _{\bar{2}}V  \notag \\
&&+2\left( \det \partial _{i}\partial _{\bar{j}}K_{(0)}\right) \left( \frac{%
y^{2}}{(Z+\frac{1}{2})_{(0)}}\partial _{y^{2}}\partial _{y^{2}}V-\partial
_{y^{2}}V\right) =0.  \label{eqn 04}
\end{eqnarray}%
This is a generalized Laplace-like equation. In the above, we have $%
V=K_{(1)},K=K_{(0)}+sK_{(1)}.$

This equation (\ref{eqn 04}) holds, starting from any smooth quarter BPS
geometry with its corresponding $K_{(0)}$, as a reference geometry, then the
linearized equation on the reference geometry in the regions outside the
enclosure surfaces is (\ref{eqn 04}). We analyze them in more details in the
following sections.

\subsection{Adding sources and superimposing sources for the function $Z$}

For the linearized equation, $Z-\frac{1}{2}$ has a superimposition
structure, whose boundary value at $y=0$, is $-1$ on the black droplets and $%
0$ on the white droplets.

Because $Z\equiv -2y^{2}\partial _{y^{2}}\partial _{y^{2}}K$, the new
solution $K\ $also induces new solution for $Z$, as
\begin{equation}
Z=Z_{(0)}+V_{(1)}.
\end{equation}%
In other words,%
\begin{equation}
Z-\frac{1}{2}=Z_{(0)}-\frac{1}{2}+V_{(1)}.
\end{equation}%
$V_{(1)}:=Z_{(1)}=-2sy^{2}\partial _{y^{2}}\partial _{y^{2}}K_{(1)}$. Hence
for the solution of the linearized equation, the boundary value of $V_{(1)}~$%
at $y=0$, is $-1$ on the black droplets and $0$ on the white droplets.

We denote $V=V_{(1)}:=Z_{(1)}$. The $Z_{(1)}$ satisfies the same form of
Laplace-like equation:
\begin{eqnarray}
&&\left( \partial _{2}\partial _{\bar{2}}K_{(0)}\right) \partial
_{1}\partial _{\bar{1}}V+\left( \partial _{1}\partial _{\bar{1}%
}K_{(0)}\right) \partial _{2}\partial _{\bar{2}}V-\left( \partial
_{1}\partial _{\bar{2}}K_{(0)}\right) \partial _{2}\partial _{\bar{1}%
}V-\left( \partial _{2}\partial _{\bar{1}}K_{(0)}\right) \partial
_{1}\partial _{\bar{2}}V  \notag \\
&&+2\left( \det \partial _{i}\partial _{\bar{j}}K_{(0)}\right) \left( \frac{%
y^{2}}{Z_{(0)}+\frac{1}{2}}\partial _{y^{2}}\partial _{y^{2}}V-\partial
_{y^{2}}V\right) =0.
\end{eqnarray}%
Then we place boundary condition at $y=0$, by adding above `$Z=-\frac{1}{2}$%
' sources as discussed in Sec. 2.1, and solve $V=Z_{(1)}\ $using the kernel
of this equation. The sources are approximated as delta function sources for
`$Z=-\frac{1}{2}$' sources. In other words, the RHS is 0 for $y>0$, and has `%
$Z=-\frac{1}{2}$' sources at $y=0$. We have the identity:
\begin{equation}
\int_{B_{i}}(-1)d^{4}z_{i}^{\prime }=(-1)v_{s,i}=-\int_{B_{i}}v_{s,i}\delta
^{(4)}(z-z_{i}^{\prime })d^{4}z_{i}^{\prime }\ .
\end{equation}%
Here $z$ denotes $(z_{1},z_{\bar{1}},z_{2},z_{\bar{2}})$. Here $i$ denotes
the $i$-th sources $\delta _{i}$. Hence
\begin{equation}
Z_{(1)}(z,0)|_{B_{i}}=-\int_{B_{i}}\delta ^{(4)}(z-z_{i}^{\prime
})d^{4}z_{i}^{\prime }=-1.
\end{equation}%
$\delta _{i}$ is at the center of $B_{i}$.

In short, we solve the equation for the kernel $f(z,y;z^{\prime })$:
\begin{eqnarray}
&&\left( \partial _{2}\partial _{\bar{2}}K_{(0)}\right) \partial
_{1}\partial _{\bar{1}}f+\left( \partial _{1}\partial _{\bar{1}%
}K_{(0)}\right) \partial _{2}\partial _{\bar{2}}f-\left( \partial
_{1}\partial _{\bar{2}}K_{(0)}\right) \partial _{2}\partial _{\bar{1}%
}f-\left( \partial _{2}\partial _{\bar{1}}K_{(0)}\right) \partial
_{1}\partial _{\bar{2}}f  \notag \\
&&+2\left( \det \partial _{i}\partial _{\bar{j}}K_{(0)}\right) \left( \frac{%
y^{2}}{Z_{(0)}+\frac{1}{2}}\partial _{y^{2}}\partial _{y^{2}}f-\partial
_{y^{2}}f\right) =0\ ;  \notag \\
&&f(z,0;z_{i}^{\prime })=\delta ^{(4)}(z-z_{i}^{\prime })  \label{kernel 02}
\end{eqnarray}%
in which the second equation is the boundary condition at the source
locations $z=z_{i}^{\prime }$. Because in linearized equation, we can put `$%
Z=-\frac{1}{2}$'\ sources in rather general locations, we can superimpose
the sources at different generic separate locations labelled by $i$. In
other words, $V_{(1)}$ has a superimposition structure for superimposing the
`$Z=-\frac{1}{2}$'\ loci.

One can obtain the kernel of this equation, which we denote as
\begin{equation}
f(z_{i},z_{\bar{j}},y;z_{i}^{\prime },z_{\bar{j}}^{\prime
}):=f(z,y;z^{\prime }),
\end{equation}%
which is a solution of (\ref{kernel 02}). Here $z$ denotes $(z_{1},z_{\bar{1}%
},z_{2},z_{\bar{2}})$. Hence
\begin{equation}
Z(z,y)=Z_{(0)}(z,y)-\sum_{i=1}^{n}\int f(z,y;z_{i}^{\prime
})d^{4}z_{i}^{\prime }.
\end{equation}%
Here $i$ denotes the $i$-th sources. The $f(z,y;z^{\prime })\ $denotes the
kernel.

We see that the above scheme is perfectly nice, since we can make both $s$
and the domain size $v_{s,i}$ of $B_{i}$ small, with a scaling relation
between them. Hence $s$ is related to the size of the enclosure surface of
each small droplet.

Denote the quanta of each small droplet as $q_{i}$. The total number of flux
quanta of small droplets are $q_{t}=\sum\limits_{i}q_{i}$. Denoting $q=%
\mathrm{max}\{q_{i}\}$, the parameter is approximately of order $s\sim
o(ql^{2}N^{-1})$. In order to linearize on the reference geometry e.g. $%
AdS_{5}\times S^{5}$, we also have $q_{i}N^{-1}\ll 1$, and $q_{i}$ can still
be large although it is much smaller than $N$. Hence $qN^{-1}\ll 1$. The
dilute distribution limit in Appendix B means $q_{i}\ll N$. This is in the
same regime as the linearized technique.

We can also consider the cases of grey droplets. Grey droplets mean that the
boundary value of $\frac{1}{2}-Z$ is not 1 but smaller than 1. These are
discussed in Sec. \ref{sec 5 3}.

\subsection{Relation between the dilute distribution regime and linearized
equation with sources}

\label{sec 2 3}

We can also consider dilute distributions of small droplets. The idea of
dilute distribution and extreme-ratio limit, and near droplet geometry has
been discussed in \cite{Chen:2007du}. We do not repeat the description here,
and we give the analysis in Appendix B. The quanta for each small droplet $%
q_{i}$ is much smaller than the total quanta $N$, i.e. $q_{i}\ll N$; and
this is referred to as the extreme-ratio limit. Dilute means that the
droplets are isolated from each other.

We refer to Sec 2.1 and 2.2 as a linearized technique. Starting from the
extreme-ratio limit, and with one central droplet much larger than other
remote small droplets, we enclose the small droplets. The region outside the
enclosures are described by the linearized equation, which are linearized on
the reference geometry. The reference geometry for the linearization, in the
simplest case, is the central big droplet. We also linearize on other smooth
droplet background, generalized from the central droplet background. The
source domain in the linearized technique, is resolved by small smooth
droplets, as in e.g. \ref{eqn 29} of Appendix B. Near each small black
droplet, that is near $y=0$,$~Z=-\frac{1}{2}$, they satisfy the smooth
droplet boundary condition.

\section{Simplification of the equations}

\renewcommand{\theequation}{3.\arabic{equation}} \setcounter{equation}{0} %
\renewcommand{\thethm}{3.\arabic{thm}} \setcounter{thm}{0} %
\renewcommand{\theprop}{3.\arabic{prop}} \setcounter{prop}{0}

\subsection{Derivation of the simplified equations}

For the asymptotic AdS space at the asymptotia, we have%
\begin{equation}
Z_{(0)}=\frac{r^{2}+y^{2}-l^{2}}{2\sqrt{(r^{2}+y^{2}-l^{2})^{2}+4y^{2}l^{2}}}%
.
\end{equation}%
Here $r^{2}=|z_{1}|^{2}+|z_{2}|^{2}$, and $l$ is a parameter (and sometimes
for simplicity are set to $l=1$ in appropriate unit). One have that $%
Z_{(0)}\equiv -2y^{2}\partial _{y^{2}}\partial _{y^{2}}K_{(0)}$. This is a
single central droplet.

Let's further make simplification of equation (\ref{kernel 02}). Now we
consider $(r^{2}+y^{2}-l^{2})^{2}\gg y^{2}l^{2}$~approximation, and
\begin{eqnarray}
Z_{(0)}-\frac{1}{2} &=&-\frac{y^{2}l^{2}}{(r^{2}+y^{2}-l^{2})^{2}}+o(l^{4})
\notag \\
&=&-\frac{y^{2}l^{2}}{(r^{2}+y^{2})^{2}}+o(l^{4}),
\end{eqnarray}%
\begin{eqnarray}
\frac{1}{Z_{(0)}+\frac{1}{2}} &=&1+\frac{y^{2}l^{2}}{(r^{2}+y^{2}-l^{2})^{2}}%
+o(l^{4})  \notag \\
&=&1+\frac{y^{2}l^{2}}{\left( r^{2}+y^{2}\right) ^{2}}+o(l^{4}).
\end{eqnarray}%
Hence%
\begin{eqnarray}
&&\left( \frac{y^{2}}{Z_{(0)}+\frac{1}{2}}\partial _{y^{2}}\partial
_{y^{2}}V-\partial _{y^{2}}V\right)  \notag \\
&=&\frac{1}{4}y^{3}\partial _{y}\frac{1}{y^{3}}\partial _{y}V+\frac{%
l^{2}y^{2}}{(r^{2}+y^{2})^{2}}y^{2}\partial _{y^{2}}\partial
_{y^{2}}V+O(l^{4}).
\end{eqnarray}

The equation (\ref{kernel 02}) is thus further simplified:
\begin{eqnarray}
&&\left( \partial _{2}\partial _{\bar{2}}K_{(0)}\right) \partial
_{1}\partial _{\bar{1}}f+\left( \partial _{1}\partial _{\bar{1}%
}K_{(0)}\right) \partial _{2}\partial _{\bar{2}}f-\left( \partial
_{1}\partial _{\bar{2}}K_{(0)}\right) \partial _{2}\partial _{\bar{1}%
}f-\left( \partial _{2}\partial _{\bar{1}}K_{(0)}\right) \partial
_{1}\partial _{\bar{2}}f  \notag \\
&&+\frac{1}{2}\left( \det \partial _{i}\partial _{\bar{j}}K_{(0)}\right)
\left( y^{3}\partial _{y}\frac{1}{y^{3}}\partial _{y}f+\frac{4l^{2}y^{2}}{%
(r^{2}+y^{2})^{2}}y^{2}\partial _{y^{2}}\partial _{y^{2}}f\right) =0\ ;
\notag \\
&&f(z,y;z_{i}^{\prime })|_{z=z_{i}^{\prime },y=\epsilon }=\delta
^{(4)}(z-z_{i}^{\prime })
\end{eqnarray}%
where
\begin{equation}
K_{(0)}=\frac{1}{2}\left( r^{2}+y^{2}\right) -\frac{1}{2}l^{2}\log \left(
r^{2}+y^{2}\right) -\frac{1}{4}y^{2}\log (y^{2}).  \label{eqn 06}
\end{equation}%
$Z_{(1)}(z,y)$ is denoted $V(z,y)$ in the above.

One may use the method of kernel which may be denoted as
\begin{equation}
f(z,y;z^{\prime }).
\end{equation}

Then we make detailed analysis of simplified equations. Denote $%
z_{1}=x_{1}+ix_{2},z_{2}=x_{3}+ix_{4}$. We have%
\begin{equation*}
\partial _{1}\partial _{\bar{1}}=\frac{1}{4}(\partial _{x_{1}}^{2}+\partial
_{x_{2}}^{2}),~~~\partial _{\bar{1}}=\frac{1}{2}(\partial _{x_{1}}+i\partial
_{x_{2}}),~\partial _{1}=\frac{1}{2}(\partial _{x_{1}}-i\partial _{x_{2}}).
\end{equation*}%
$l^{2}$ is a parameter in this class of solutions. Here, we solve the
linearized equation up to order $O(l^{2})$. Denoting $%
T=|z_{1}|^{2}+|z_{2}|^{2}+y^{2}$,
\begin{eqnarray}
\partial _{2}\partial _{\bar{2}}K_{(0)} &=&1/2-\frac{l^{2}}{2}\frac{1}{T}+%
\frac{l^{2}}{2}\frac{|z_{2}|^{2}}{T^{2}},~~\partial _{1}\partial _{\bar{1}%
}K_{(0)}=1/2-\frac{l^{2}}{2}\frac{1}{T}+\frac{l^{2}}{2}\frac{|z_{2}|^{2}}{%
T^{2}},  \notag \\
\partial _{1}\partial _{\bar{2}}K_{(0)} &=&\frac{l^{2}}{2}\frac{\bar{z}%
_{1}z_{2}}{T^{2}},~~~~~~~~\partial _{2}\partial _{\bar{1}}K_{(0)}=\frac{l^{2}%
}{2}\frac{z_{1}\bar{z}_{2}}{T^{2}},  \notag \\
\det \partial _{i}\partial _{\bar{j}}K_{(0)} &=&1/4-\frac{l^{2}}{2T}+\frac{%
l^{2}(|z_{1}|^{2}+|z_{2}|^{2})}{4T^{2}}.
\end{eqnarray}%
Hence,
\begin{eqnarray}
&&\left( \partial _{2}\partial _{\bar{2}}K_{(0)}\right) \partial
_{1}\partial _{\bar{1}}V+\left( \partial _{1}\partial _{\bar{1}%
}K_{(0)}\right) \partial _{2}\partial _{\bar{2}}V-\left( \partial
_{1}\partial _{\bar{2}}K_{(0)}\right) \partial _{2}\partial _{\bar{1}%
}V-\left( \partial _{2}\partial _{\bar{1}}K_{(0)}\right) \partial
_{1}\partial _{\bar{2}}V  \notag \\
&&+\frac{1}{2}\left( \det \partial _{i}\partial _{\bar{j}}K_{(0)}\right)
\left( y^{3}\partial _{y}\frac{1}{y^{3}}\partial _{y}V+\frac{4l^{2}y^{2}}{%
(r^{2}+y^{2})^{2}}y^{2}\partial _{y^{2}}\partial _{y^{2}}V\right)  \notag \\
&=&(D_{0}+D_{2})V,
\end{eqnarray}%
where
\begin{eqnarray}
D_{0} &=&\frac{1}{2}\partial _{1}\partial _{\bar{1}}+\frac{1}{2}\partial
_{2}\partial _{\bar{2}}+\frac{1}{8}\left( y^{3}\partial _{y}\frac{1}{y^{3}}%
\partial _{y}\right)  \notag \\
&=&\frac{1}{8}(\sum_{i=1}^{4}\partial _{x_{i}}^{2}+y^{3}\partial _{y}\frac{1%
}{y^{3}}\partial _{y}),
\end{eqnarray}%
and
\begin{align}
D_{2}& =-\frac{l^{2}}{2}\frac{\bar{z}_{1}z_{2}}{T^{2}}\partial _{\bar{z}%
_{1}}\partial _{z_{2}}-\frac{l^{2}}{2}\frac{z_{1}\bar{z}_{2}}{T^{2}}\partial
_{z_{1}}\partial _{\bar{z}_{2}}+(-\frac{l^{2}}{2}\frac{1}{T}+\frac{l^{2}}{2}%
\frac{|z_{2}|^{2}}{T^{2}})\partial _{z_{1}}\partial _{\bar{z}_{1}}+(-\frac{%
l^{2}}{2}\frac{1}{T}+\frac{l^{2}}{2}\frac{|z_{1}|^{2}}{T^{2}})\partial
_{z_{2}}\partial _{\bar{z}_{2}}  \notag \\
& +(-\frac{l^{2}}{2T}+\frac{l^{2}(|z_{1}|^{2}+|z_{2}|^{2})}{4T^{2}})\frac{1}{%
2}y^{3}\partial _{y}\frac{1}{y^{3}}\partial _{y}+\frac{l^{2}y^{4}}{2T^{2}}%
\partial _{y^{2}}\partial _{y^{2}},
\end{align}%
and $T=|z_{1}|^{2}+|z_{2}|^{2}+y^{2}.$

We write $f=f_{0}+f_{2}$, and the solution is of the form
\begin{equation}
f(z,y;z^{\prime })=\frac{cy^{4}}{\left( |z_{1}-z_{1}^{\prime
}|^{2}+|z_{2}-z_{2}^{\prime }|^{2}+y^{2}\right) ^{4}}+f_{2}.  \label{eqn 08}
\end{equation}%
Near $y=0$, $f$ approaches 4d delta function, of 4d coordinates $%
(z-z^{\prime })$, with some coefficient. Finite small $y^{2}$ plays the role
of smoothing of the delta function.

In the below section, we can also perform a regularization, to place the
delta function at $y=\epsilon $.

We have%
\begin{equation}
Z\left( z,y\right) -\frac{1}{2}=Z_{(0)}\left( z,y\right) -\frac{1}{2}%
-\int_{\cup _{i}B_{i}}f(z,y;z^{\prime })d^{4}z^{\prime }~.
\end{equation}%
In other words, the solution $Z\left( z,y\right) $ to the linearized
equation is in terms of the integral of the kernel $f(z,y;z^{\prime })$. In
order to get the solution for $Z\left( z,y\right) $, we need to solve $%
f(z,y;z^{\prime })$, which is a function of $l^{2}$. We integrate over
domain denoted $B_{i}$, to smooth-out and straighten the distribution
labeled by $i$. The $f(z,y;z^{\prime })$ is a nontrivial function of $l^{2}$%
, since the Laplace-type equation contains $l^{2}$ as a parameter.

\subsection{Summary of equations and boundary conditions}

The differential operator $D$ is
\begin{equation}
D=D_{0}+D_{2},
\end{equation}%
and $D=D(l^{2})$ is a function of $l^{2}.$
\begin{equation}
D_{0}=\frac{1}{8}(\sum_{i=1}^{4}\partial _{x_{i}}^{2}+y^{3}\partial _{y}%
\frac{1}{y^{3}}\partial _{y}),
\end{equation}%
\begin{align}
D_{2}& =-\frac{l^{2}}{2}\frac{\bar{z}_{1}z_{2}}{T^{2}}\partial _{\bar{z}%
_{1}}\partial _{z_{2}}-\frac{l^{2}}{2}\frac{z_{1}\bar{z}_{2}}{T^{2}}\partial
_{z_{1}}\partial _{\bar{z}_{2}}+(-\frac{l^{2}}{2}\frac{1}{T}+\frac{l^{2}}{2}%
\frac{|z_{2}|^{2}}{T^{2}})\partial _{z_{1}}\partial _{\bar{z}_{1}}+(-\frac{%
l^{2}}{2}\frac{1}{T}+\frac{l^{2}}{2}\frac{|z_{1}|^{2}}{T^{2}})\partial
_{z_{2}}\partial _{\bar{z}_{2}}  \notag \\
& +(-\frac{l^{2}}{2T}+\frac{l^{2}(|z_{1}|^{2}+|z_{2}|^{2})}{4T^{2}})\frac{1}{%
2}y^{3}\partial _{y}\frac{1}{y^{3}}\partial _{y}+\frac{l^{2}y^{4}}{2T^{2}}%
\partial _{y^{2}}\partial _{y^{2}}.
\end{align}

The kernel $f=f(z,y;z^{\prime })$ is
\begin{equation}
f(z,y;z^{\prime })=f_{0}(z,y;z^{\prime })+f_{2}(z,y;z^{\prime }),
\end{equation}%
where $f_{0}$ is of the form in (\ref{eqn 08}). Here, $y=0$ is the boundary
and $y>0$ is the bulk.

We can perform a regularization, by considering $y=\epsilon $ instead of $%
y=0 $, where $\epsilon $ is a small regularization parameter.

These are the equation and boundary behavior near $y=\epsilon $.$~$Since $%
D=D(l^{2})$ is a function of $l^{2}$, $f$ is also a function of $l^{2}$. The
equation can be written as
\begin{eqnarray}
(D_{0}+D_{2})(f_{0}+f_{2}) &=&0,~~~~y>\epsilon , \\
\left( f_{0}+f_{2}\right) (z,y;z^{\prime })|_{y=\epsilon } &=&\delta
^{(4)}(z-z^{\prime }).
\end{eqnarray}%
We will solve $f=f_{0}+f_{2}$. The equation is simplified to
\begin{eqnarray}
&&D_{0}f_{0}=0,~D_{0}f_{2}+D_{2}f_{0}=0,~~~y>\epsilon ,  \label{eqn 17} \\
&&\left( f_{0}+f_{2}\right) (z,y;z^{\prime })|_{y=\epsilon }=\delta
^{(4)}(z-z^{\prime }).  \label{eqn 18}
\end{eqnarray}

Hence the equation for $f_{2}$ is
\begin{equation}
\frac{1}{8}(\sum_{i=1}^{4}\partial _{x_{i}}^{2}+y^{3}\partial _{y}\frac{1}{%
y^{3}}\partial _{y})f_{2}=-D_{2}\left( \frac{cy^{4}}{\left(
|z_{1}-z_{1}^{\prime }|^{2}+|z_{2}-z_{2}^{\prime }|^{2}+y^{2}\right) ^{4}}%
\right) .  \label{eqn 14}
\end{equation}
The behavior of $f=f_{0}+f_{2}$ near $y=\epsilon $ is, in general,
\begin{equation}
\lim_{y\rightarrow \epsilon }f(z,y;z^{\prime })=\delta ^{(4)}(z-z^{\prime }).
\end{equation}

In the later sections, we will discuss both the cases when the delta
function behavior is placed at $y=0$ and at $y=\epsilon $ respectively.

\section{Equations in Fourier space and derivation of global solutions}

\renewcommand{\theequation}{4.\arabic{equation}} \setcounter{equation}{0} %
\renewcommand{\thethm}{4.\arabic{thm}} \setcounter{thm}{0} %
\renewcommand{\theprop}{4.\arabic{prop}} \setcounter{prop}{0}

In this section, we analyze the general equation for the function $f$, and
give solutions for asymptotic AdS backgrounds. The solutions come with
parameters which are related to boundary conditions and asymptotia.

Denote the momentum variable dual to $x_{i}$ by $k_{i}$, then $%
\sum_{i=1}^{4}\partial _{x_{i}}^{2}$ is turned into multiplication by $%
-\sum_{i=1}^{4}k_{i}^{2}$, and we denote it as $-k^{2}$. We denote $%
s_{i}=x_{i}-x_{i}^{\prime }$, and the norm of $s_{i}$ is denoted $s$. The
norm of $k_{i}$ is denoted $k$.

We denote $\mu =y^{2}\ $and we have
\begin{equation}
D_{2}f_{0}=D_{2}(\frac{cy^{4}}{(s^{2}+y^{2})^{4}})=-cl^{2}\mu ^{2}(\mu
+s^{2})^{-6}.
\end{equation}%
We have
\begin{equation}
\frac{1}{8}(\sum_{i=1}^{4}\partial _{x_{i}}^{2}+4\mu \partial _{\mu
}^{2}-4\partial _{\mu })f_{2}=-D_{2}(\frac{cy^{4}}{(|z_{1}-z_{1}^{\prime
}|^{2}+|z_{2}-z_{2}^{\prime }|^{2}+y^{2})^{4}}).
\end{equation}

We can perform radial Fourier transforms in 4d. The $k$ is the conjugate
variable to $s$. Denote the radial Fourier transform of $f(s,\mu )$ as
\begin{equation}
F[f](k,\mu )=F(k,\mu ).
\end{equation}%
Then
\begin{equation}
f=\int \frac{d^{4}k}{(2\pi )^{4}}F(k,\mu )e^{ik\cdot x},  \label{eqn 26}
\end{equation}%
\begin{equation}
F(k,\mu )=\int d^{4}xf(s,\mu )e^{-ik\cdot x}.
\end{equation}%
We denote $F_{0}(k,\mu ):=F[f_{0}](k,\mu )$ and $F_{2}(k,\mu
):=F[f_{2}](k,\mu )$. The inverse Fourier transform is denoted $F^{-1}[\cdot
].$

According to Hankel transform and its inverse Hankel transform,
\begin{equation}
kF(k,\mu )=(2\pi )^{2}\int_{0}^{\infty }J_{1}(ks)sf(s,\mu )sds;
\end{equation}%
\
\begin{equation}
sf(s,\mu )=(2\pi )^{2}\int_{0}^{\infty }J_{1}(ks)kF(k,\mu )kdk.
\end{equation}

From above, we have
\begin{equation}
(\partial _{s}^{2}+3s^{-1}\partial _{s}+4\mu \partial _{\mu }^{2}-4\partial
_{\mu })f_{2}=8cl^{2}\mu ^{2}(\mu +s^{2})^{-6}.  \label{eqn 30}
\end{equation}%
The $f_{0}$ satisfies the equation
\begin{equation}
(\partial _{s}^{2}+3s^{-1}\partial _{s}+4\mu \partial _{\mu }^{2}-4\partial
_{\mu })f_{0}=0,~~~~y>0.
\end{equation}%
Hence we have,
\begin{equation}
(\partial _{s}^{2}+3s^{-1}\partial _{s}+4\mu \partial _{\mu }^{2}-4\partial
_{\mu })f=8cl^{2}\mu ^{2}(\mu +s^{2})^{-6},~\   \label{eqn 32}
\end{equation}%
where $f=f_{0}+f_{2}$.$\ $The $f_{0}$ is the piece annihilated by the
differential operator on the left-hand side; while $f_{2}$ when acted by the
differential operator, has a source term on the right-hand side.

We consider both the cases, when the boundary condition is at $y=0$; and
when the boundary condition is at $y=\epsilon $. The latter can be
considered as a regularization of the former.

We can obtain the Fourier transform for the equation for $f(s,y)$. Multiply
both sides of (\ref{eqn 32}) by $\mu $:
\begin{equation}
\mu (\partial _{s}^{2}+3s^{-1}\partial _{s}+4\mu \partial _{\mu
}^{2}-4\partial _{\mu })f=8cl^{2}\mu ^{3}(\mu +s^{2})^{-6}.
\end{equation}%
We have the Fourier transform $F[\frac{y^{6}}{\left( y^{2}+s^{2}\right) ^{6}}%
]=\xi y^{2}k^{4}K_{4}(ky)$, where $\xi =\frac{4\pi ^{2}}{3840}$. We also have%
\begin{equation}
F\left[ \mu (\partial _{s}^{2}+3s^{-1}\partial _{s}+4\mu \partial _{\mu
}^{2}-4\partial _{\mu })\right] =-\mu k^{2}+4\mu ^{2}\partial _{\mu
}^{2}-4\mu \partial _{\mu }\ ,
\end{equation}%
\begin{equation}
F[f(s,\mu )]=F[f](k,\mu )=F(k,\mu )\ .
\end{equation}%
So we have the momentum space version of the equation (\ref{eqn 32}):%
\begin{equation}
\left( -y^{2}k^{2}+y^{2}\partial _{y}^{2}-3y\partial _{y}\right) F(k,y)=8\xi
cy^{2}l^{2}k^{4}K_{4}(ky)  \label{eqn 40}
\end{equation}%
since%
\begin{equation}
4\mu \partial _{\mu }^{2}-4\partial _{\mu }=\partial
_{y}^{2}-3y^{-1}\partial _{y}\ .
\end{equation}%
This equation does not assume using the regularization at $y=\epsilon $. For
simplicity we may also denote $\epsilon $ as $y_{c}$ in this section. The
equation itself does not know about the limiting parameter $\epsilon $.

We can solve the solution by expansion using modified Bessel functions $%
K_{n}(ky)$. Note
\begin{equation}
\left( -\mu k^{2}+4\mu ^{2}\partial _{\mu }^{2}-4\mu \partial _{\mu }\right)
\frac{y^{2}K_{2}(ky)}{K_{2}(ky_{c})}=0.  \label{eqn 39}
\end{equation}%
We get
\begin{equation}
F(k,y)=F_{0}(k,y)+F_{2}(k,y)=c_{1}\frac{y^{2}K_{2}(ky)}{%
y_{c}^{2}K_{2}(ky_{c})}+c_{2}\frac{y^{2}l^{2}K_{4}(ky)}{%
y_{c}^{4}K_{4}(ky_{c})}.  \label{eqn 42}
\end{equation}%
Then we analyze more details of the coefficients $c_{1},c_{2}$.

Note that the modified Bessel functions $K_{n}(ky)$ themselves have
recurrence relations
\begin{equation}
K_{n+1}(ky)-K_{n-1}(ky)=\frac{2n}{ky}K_{n}(ky).
\end{equation}%
Note we do not consider $I_{n}(ky)$ since for large $y$, $I_{n}(ky)\sim
e^{ky}/\sqrt{ky}$, while $K_{n}(ky)\sim e^{-ky}/\sqrt{ky}$. They have the
properties:
\begin{equation}
\mathrm{when}\ y_{c}\rightarrow 0,\ \ K_{n}(ky_{c})\rightarrow \frac{\Gamma
(n)}{2}(\frac{2}{ky_{c}})^{n},
\end{equation}%
e.g.
\begin{equation}
\mathrm{when}~y_{c}\rightarrow 0,~\frac{1}{y_{c}^{4}K_{4}(ky_{c})}%
\rightarrow \frac{k^{4}}{48},
\end{equation}%
\begin{equation}
\frac{1}{y_{c}^{2}K_{2}(ky_{c})}\rightarrow \frac{k^{2}}{2}.
\end{equation}%
The above (\ref{eqn 40}) implies
\begin{equation}
F_{2}(k,y)=\frac{2}{3}\xi cl^{2}y^{2}k^{4}K_{4}(ky)\simeq 32\xi c\frac{%
y^{2}l^{2}K_{4}(ky)}{y_{c}^{4}K_{4}(ky_{c})}.
\end{equation}%
Here the $\simeq $ sign means \textquotedblleft equal when $y_{c}\rightarrow
0$\textquotedblright . The (\ref{eqn 39}) implies
\begin{equation}
F_{0}(k,y)=\frac{c_{1}}{2}y^{2}k^{2}K_{2}(ky)\simeq c_{1}\frac{y^{2}K_{2}(ky)%
}{y_{c}^{2}K_{2}(ky_{c})}.
\end{equation}%
In the regularization case, we have families of solutions with parameters $%
y_{c}>0$, and since $y_{c}$ is a cut-off, it is hence tunable. We can also
take$~y_{c}\rightarrow 0^{+}$ limit.

The combination\ $F_{0}(k,y)+F_{2}(k,y)\ $approaches 1, when $y$ approaches $%
y_{c}$. This means that $c_{1}+\frac{c_{2}l^{2}}{y_{c}^{2}}=1$. The
condition corresponds to the boundary value of $\frac{1}{2}-Z(z,y)$ to be 1,
at the black regions in the interior end, which originates from the smooth
droplet boundary conditions.

We can equivalently write:
\begin{equation}
F(k,y)\simeq c_{1}\frac{y^{2}K_{2}(ky)}{y_{c}^{2}K_{2}(ky_{c})}+c_{2}\frac{%
y^{2}l^{2}K_{4}(ky)}{y_{c}^{4}K_{4}(ky_{c})},\ \ \
\end{equation}
\begin{equation}
F(k,y)\simeq c_{1}\frac{1}{2}k^{2}y^{2}K_{2}(ky)+c_{2}\frac{1}{48}%
k^{4}l^{2}y^{2}K_{4}(ky),\ \   \label{eqn 54}
\end{equation}%
where $c_{1}=1-c_{2}\frac{l^{2}}{y_{c}^{2}}$.

The function $f$ may be written as
\begin{eqnarray}
f &=&f_{0}+f_{2}=F^{-1}[F[f]]  \notag \\
&\simeq &\frac{y^{2}}{y_{c}^{2}}\int \frac{d^{4}k}{(2\pi )^{4}}\left( \frac{%
c_{1}K_{2}(ky)}{K_{2}(ky_{c})}+\frac{c_{2}l^{2}K_{4}(ky)}{%
y_{c}^{2}K_{4}(ky_{c})}\right) e^{ik\cdot (z-z^{\prime })}.
\end{eqnarray}

\section{Stretched technique and corresponding solutions in Fourier space}

\label{sec 5} \renewcommand{\theequation}{5.\arabic{equation}} %
\setcounter{equation}{0} \renewcommand{\thethm}{5.\arabic{thm}} %
\setcounter{thm}{0} \renewcommand{\theprop}{5.\arabic{prop}} %
\setcounter{prop}{0}

\subsection{General methods of stretched technique}

Here we consider another solution technique, the stretched surface
technique. This is similar to the idea in \cite{Balasubramanian:2018yjq}.
The stretched technique can be naturally combined with the technique in
sections 2--4. The linearized technique put small enclosure surfaces near
each small droplets. The stretched technique, put a single surface at $%
y=y_{c}$. The small enclosure surfaces are all inside the stretched surface.
The stretched surface can hence be viewed as a big and universal enclosure
surface.

In the stretched surface limit, one can push-forward the droplet
distributions onto the stretched surface at $y=y_{c}$. The change with
respective to the $y_{c}$ variable is analogous to renormalization group
transformation. We transport the data from a smaller $y$ surface to a bigger
$y$ surface, and this procedure is a coarse-graining over the high-momentum
modes on the stretched surface. All the droplets at $y=0$ are enclosed by $%
y=y_{c}$ stretched surface, as a big enclosure surface.

When we combine both techniques, the solutions are asymptotically AdS space,
whose asymptotic behaviors are (\ref{eqn 06}), hence a coarse-graining
description of them also describes superstar and their deformations. The
boundary condition for superstars is a coarse-grained density and hence a
grey droplet density, see Sec. 5.3.1.

For asymptotic AdS background, we place an interior stretched surface at $%
y_{c}$, and the equation is
\begin{equation}
\left( -y^{2}k^{2}+y^{2}\partial _{y}^{2}-3y\partial _{y}\right) F(k,y)=%
\frac{c_{2}}{4}l^{2}y^{2}k^{4}K_{4}(ky).
\end{equation}%
For asymptotic AdS,
\begin{equation}
F(k,y)\simeq c_{1}\frac{y^{2}K_{2}(ky)}{y_{c}^{2}K_{2}(ky_{c})}+c_{2}\frac{%
y^{2}l^{2}K_{4}(ky)}{y_{c}^{4}K_{4}(ky_{c})},\ \ \ \ y\geq y_{c}
\end{equation}%
and $c_{1}+c_{2}\frac{l^{2}}{y_{c}^{2}}=\frac{1}{1+\nu }$. Hence,
\begin{eqnarray}
c_{1} &=&\frac{1}{1+\nu }-c_{2}\frac{l^{2}}{y_{c}^{2}}, \\
c_{2}/c_{1} &=&\gamma \quad =\quad \frac{\frac{1}{1+\nu }-c_{1}}{c_{1}}\frac{%
y_{c}^{2}}{l^{2}},
\end{eqnarray}%
where $0\leq \frac{1}{1+\nu }\leq 1$. The solutions have parameters $y_{c}$,
$l$, $c_{1}$ and $\nu $. When $y_{c}^{2}$ is increased, $\gamma $ tends to
become larger. For coarse-grained grey droplet, this means that the boundary
value is not 1, but $\frac{1}{1+\nu }$. The condition corresponds to the
boundary value of $\frac{1}{2}-Z(z,y)$ to be $\frac{1}{1+\nu }$, at the
black regions in the interior end. This describes situations of superstar
geometries, see Sec. \ref{sec 5 3}. Note $\nu =0$ is a special case, where
the charge of superstar goes to zero. The function contains parameter $y_{c}$%
, hence we also denote it as $F(k,y)_{y_{c}}$.

Now the data on $y=y_{c}$ corresponds to a grey distribution, and they
correspond to coarse-grained microstates of two-charge superstar. This idea
is similar to Sec. 4 and 5 of \cite{Balasubramanian:2018yjq}. Hence the
boundary data is $\frac{1}{1+\nu }$ instead of 1, and see the analysis in
Sec. \ref{sec 5 3}. The $y=y_{c}$ can be viewed as a stretched surface,
where the droplet distribution data are pushed-forward from the interior to
this stretched surface, and hence the droplet data is coarse-grained on the
stretched surface. The $l$ is a parameter for the asymptotic AdS radius, and
$\nu $ is a parameter related to the total charge of the superstar.

The smooth geometries correspond to smooth droplet boundary conditions in
the interior, i.e. $y=0$. Now we consider a distribution of droplets,
resulting in a coarse distribution in the $y=y_{c}$ stretched droplet space.
The $z$ are four-dimensional coordinates. We define $u(z,y)=\frac{1}{2}%
-Z(z,y)$. We place a coarse distribution at $y=y_{c}$ stretched droplet
space, and hence $u(z,y)|_{y=y_{c}}=u_{0}(z,y_{c})$ is the coarse
distribution function at $y=y_{c}$ stretched droplet space. We use the
convolution with the kernel to push-forward this distribution to the bulk.
We refer to this stretched droplet space limit, as \textquotedblleft
stretched limit". Hence
\begin{equation}
u(z,y)=\int_{\cup _{i}C_{i}}f(z,z^{\prime },y)_{y_{c}}u_{0}(z^{\prime
},y_{c})d^{4}z^{\prime },
\end{equation}%
where $f(z,z^{\prime },y)_{y_{c}}=F^{-1}[F(k,y)_{y_{c}}]$. Here $%
u_{0}(z^{\prime },y_{c})$ includes the distribution of all the droplets. The
Fourier transform, as well as Mellin transform, has nice convolution
properties,%
\begin{equation}
F[u(z,y)]=F(k,y)F[u_{0}(z,y_{c})].
\end{equation}%
The metric function by the stretched droplet space can be written as, in the
stretched limit
\begin{equation}
Z\left( z,y\right) -\frac{1}{2}=-\int_{\cup _{i}C_{i}}f(z,z^{\prime
},y)_{y_{c}}u_{0}(z^{\prime },y_{c})d^{4}z^{\prime }~.
\end{equation}%
The $C_{i}$ are domains on $y=y_{c}$ stretched surface. Here $%
u_{0}(z^{\prime },y_{c})$ has included the distribution (denoted $C_{0}$ in $%
\cup _{i}C_{i}$) of the central droplet pushed-forward to the $y=y_{c}$
stretched surface. The stretched surface includes all the droplets inside.
The Fourier transform of the metric function can be written as in the
stretched limit
\begin{equation}
F\left[ \frac{1}{2}-Z\left( z,y\right) \right] (k,y)=F(k,y)F[u_{0}(z,y_{c})].
\end{equation}%
Note $y_{c}$ is a parameter. Changing the $y_{c}$ infinitesimally is
changing the scale of coarse-graining and is analogous to renormalization
group transformation.

Hence the Fourier transform of the metric function can be written as, in the
stretched limit%
\begin{equation}
F\left[ \frac{1}{2}-Z\left( z,y\right) \right] (k,y)=\frac{y^{2}}{y_{c}^{2}}%
\left( \frac{c_{1}K_{2}(ky)}{K_{2}(ky_{c})}+\frac{c_{2}l^{2}K_{4}(ky)}{%
y_{c}^{2}K_{4}(ky_{c})}\right) F[u_{0}(z,y_{c})].
\end{equation}%
The metric function by the stretched droplet space can be written as, in the
stretched limit
\begin{eqnarray}
Z\left( z,y\right) -\frac{1}{2} &=&-\int_{\cup _{i}C_{i}}f(z,z^{\prime
},y)_{y_{c}}u_{0}(z^{\prime },y_{c})d^{4}z^{\prime }~  \notag \\
&\simeq &-\int_{\cup _{i}C_{i}}\int \frac{d^{4}k}{(2\pi )^{4}}\frac{y^{2}}{%
y_{c}^{2}}\left( \frac{c_{1}K_{2}(ky)}{K_{2}(ky_{c})}+\frac{%
c_{2}l^{2}K_{4}(ky)}{y_{c}^{2}K_{4}(ky_{c})}\right) e^{ik\cdot (z-z^{\prime
})}u_{0}(z^{\prime },y_{c})d^{4}z^{\prime }.  \notag \\
&&  \label{eqn 66}
\end{eqnarray}%
These cases (\ref{eqn 66}) are in the stretched limit, where the boundary
condition is placed at the stretched surface with a density $u_{0}(z^{\prime
},y_{c})$. The droplets on the stretched surface are grey droplets, whose
distribution are described by $u_{0}(z^{\prime },y_{c}).$

The above stretched limit (\ref{eqn 66}), is naturally used for the
superstar, which has a natural cut-off \cite{Balasubramanian:2018yjq}. One
can get coarse graining of distributions for the two-charge superstar. We
place the above stretched surface at $y=y_{0}$ for superstar, and this is
similar to \cite{Balasubramanian:2018yjq}. The coarse graining of these
smooth geometries leads to e.g. the superstar geometry, which is an
effective geometry. Low energy observers see an averaged metric, which is an
effective geometry.

We can also perform coarse-graining at a coarse-graining scale $k_{0}\sim
\frac{1}{y_{0}}$, i.e. at $y=y_{0}$. For coarse-graining and renormalization
group flow purpose, we use a coarse-graining function $P_{IR}(k)$,
\begin{equation}
P_{IR}(k)\simeq F(k,y_{0})\simeq c_{1}\frac{1}{2}%
k^{2}y_{0}^{2}K_{2}(ky_{0})+c_{2}\frac{1}{48}k^{4}l^{2}y_{0}^{2}K_{4}(ky_{0})
\label{eqn 62}
\end{equation}%
which is a smooth filter reducing the UV modes higher than scale $y_{0}^{-1}$%
. This is a momentum filter function, as similar to \cite%
{Balasubramanian:2018yjq}. We can transport, in other words, push-forward
the data from $y=0$ outwards to $y=y_{0}$ surface. See also Appendix E and (%
\ref{eqn 71}), for the use of $P_{IR}(k)$ in the coarse-graining.

\subsection{Relation between linearized technique, regularization technique,
and stretched technique}

\label{sec 5 2}

The linearized technique is, a priori, independent of regularization
technique, and stretched limit. It can be naturally combined with
regularization technique and with stretched technique.

The regularization technique is encountered when we place boundary
conditions at $y=\epsilon $. It is a regularization for the case $\epsilon $
goes to zero. The boundary conditions are those originate from smooth
droplet boundary conditions.

The stretched technique, is different from the regularization technique. The
two can be naturally combined. The stretched technique is placing $y=y_{c}$
stretched surface. The boundary conditions are more general than the
regularization technique. We can place grey droplet boundary conditions.
Among other things, these include two-charge superstar spacetimes and other
spacetimes including spacetime foams, e.g. related to \cite%
{Jejjala:2008jy,Caldarelli:2004mz,Berenstein:2007wz,Balasubramanian:2005mg}.
Changing $y_{c}$ is analogous to renormalization group transformation; and
we can also see this from the combination of $y_{c}k$, e.g. scaling up $%
y_{c} $ is related to scaling smaller $k$. The bulk geometry, automatically
and efficiently, geometrized this renormalization group process.

\subsection{Relation to superstar}

\label{sec 5 3}

\subsubsection{Relation to superstar and general aspects}

\label{sec 5 3 1}

Here we analyze the coarse density for superstar for the boundary condition.
The asymptotics of superstar are analyzed in Appendix C, where we have shown
that the dilute distribution of small droplets, in the asymptotic region
matches with the asymptotic expansion of the superstar. Hence we consider a
coarse-grained description.

Consider a stretched surface or stretched horizon for superstar, at $y=y_{0}$%
. The density $u_{0}(z,y_{0})$ defines a coarse distribution, and this
defines also a coarse-grained distribution from microstates. It has a
distribution on the stretched surface. The distribution is related to Young
tableau (YT) profiles.

The two-charge superstars have also been analyzed in \cite{Chen:2007du}, and
the $Z$ function for superstar is
\begin{eqnarray}
\frac{1}{2}-Z &=&\frac{1}{1+Q_{2}+Q_{3}+{\tilde{r}}^{2}+{Q_{2}\,Q_{3}}/{%
\tilde{r}^{2}}+\cot ^{2}\theta \lbrack ({\tilde{r}}^{2}+Q_{3})\sin
^{2}\alpha +({\tilde{r}}^{2}+Q_{2})\cos ^{2}\alpha ]}\ .  \notag \\
&&
\end{eqnarray}%
We work in unit $l=1$. Note the \textquotedblleft $1$\textquotedblright\ on
the right hand side correspond to $l^{2}$, when restored. Consider the
identification \cite{Chen:2007du} that $y={\tilde{r}}\sin \theta $,$~r={%
\tilde{r}}\cos \theta $, and $r^{2}={\tilde{r}}^{2}-y^{2}$, and~if $y$ is
small, then the density is concentrated in the small $r$ region, due to the $%
\cot ^{2}\theta $ term. When close to small ${\tilde{r}}^{2}$ region, the
superstar geometry is replaced by smooth microstate geometries; there we use
a small $y^{2}$ expansion to extract the droplet distributions, e.g. (\ref%
{eqn 67}).

Hence we assume the density is within $|z_{1}|^{2}\leqslant r_{1,0}^{2}$, $%
|z_{2}|^{2}\leqslant r_{2,0}^{2}$. The coarse density corresponds to the
coarse-grained density of droplets for superstars. The fine density is the
distribution of smooth droplet configurations of the smooth microstate
geometries. Here, for superstar we consider the coarse density. For the
moment, we first assume small charge regimes $Q_{i}\ll l^{2}$.

Note that in the half BPS case, if one redefines the function $%
-z(x,y)\rightarrow z(x,y)$ there \cite{Lin:2004nb}, the system is the same.
Here we have chosen a convention to use, and here we use $Z(z,y)$ instead of
using $-Z(z,y)$. With this convention, $0\leq (\frac{1}{2}-Z)_{y_{0}}(z)\leq
1$,%
\begin{equation}
u(z)_{y_{0}}=(\frac{1}{2}-Z)_{y_{0}}(z).
\end{equation}%
Then we simplify%
\begin{eqnarray}
u(z)_{y_{0}} &=&(\frac{1}{2}-Z)|_{y=y_{0}}(z)\approx \frac{1}{1+Q_{2}+Q_{3}}
\label{eqn 55} \\
&\simeq &\frac{1}{1+\nu },  \label{eqn 55 02}
\end{eqnarray}%
where $\nu $ is related to the total charge of superstar. We can have a
small ${Q_{3}}$ limit. The small ${Q_{3}}$ limit of (\ref{eqn 55}) goes over
to the results similar to \cite{Balasubramanian:2005mg} and their related
aspects. Among other things, mass, radius and charge relations can be
produced by entropic maximization related methods \cite%
{Balasubramanian:2005mg}. See also discussions on related aspects, e.g. \cite%
{Balasubramanian:2007bs,Fareghbal:2008ar}. Grey droplets refer to that the
boundary value of $\frac{1}{2}-Z$ is not 1 but smaller than 1. The
coarse-grained distributions are grey droplets. Here, $\nu $ is related to
the slopes of YTs.

One can describe the subsystem here by two chemical potentials and an
effective canonical temperature, in canonical ensembles, microcanonical
ensembles and other related ensembles depending on changes of variables and
Legendre transformations \cite%
{Balasubramanian:2005mg,Balasubramanian:2018yjq}. The two chemical
potentials conjugate to the row lengths of two YTs $r_{1},r_{2}~$for the $Y$
and $X$ fields. We also refer to these two Young tableaux as YT$_{1}$ and YT$%
_{2}$. The YTs of this kind have been constructed in e.g. \cite%
{Kimura:2007wy} and their related references \cite{Kimura:2011df}. These YTs
can also be seen from the charge relation. One of the simplest cases are
operators of the form in e.g. \cite{Kimura:2011df}. See also related
construction \cite{Lewis-Brown:2020nmg,Bhattacharyya:2008rb}\footnote{%
The shapes of Young tableaux have been considered in e.g. \cite%
{Balasubramanian:2005mg,Balasubramanian:2018yjq,Simon:2018laf,Berenstein:2023srv,Ahmadain:2022gfw}
and their related references.}. These constructions are equivalent and
related to each other, by a change of basis. In the next subsection \ref{sec
5 3 2}, we also consider a similar coherent state version of the YT states.

Negative ${Q_{3}}${\ or }${Q_{2}}${,} can make{\ }$(\frac{1}{2}%
-Z)_{y_{0}}(z)>1$, where there are closed time-like curves. The case of
negative charges has been discussed in e.g. \cite%
{Myers:2001aq,Kimura:2007wy,Caldarelli:2004mz,Kimura:2011df} and their
related references and aspects.

We hence have a 4d density $u_{0}(z)$ on the stretched surface $y=y_{0}$,
and one transports $u_{0}(z,y_{0})$ to $u(z,y)$ at a larger radial location $%
y$. This idea has been considered in \cite{Balasubramanian:2018yjq}; and the
set-up here is a generalization. We have
\begin{equation}
u\left( z,y\right) =\int_{C}f(z,z^{\prime },y)_{y_{0}}u_{0}(z^{\prime
},y_{0})d^{4}z^{\prime }\ ,
\end{equation}%
where $f(z,z^{\prime },y)_{y_{0}}=F^{-1}[F(k,y)_{y_{0}}]$ and $C$ is the
domain where $u_{0}\simeq (1+\nu )^{-1}$. For simplicity, consider equal
charge case, $C$ is $r^{2}\leq r_{0}^{2}$,$\ $where $r_{0}^{2}\simeq
l^{2}(1+\nu )$. One can generate deformations around superstar by deforming
the density from $u_{0}(z,y_{0})$ to
\begin{equation}
u_{0}(z,y_{0})+\delta u_{0}(z,y_{0}),
\end{equation}%
which is a general distribution with general variations. This deformed
density is transported to a larger radial position $y$. See also appendix E
and (\ref{eqn 71}), for the use of coarse-graining procedure.

\subsubsection{Additional discussions and relation to the symplectic form}

\label{sec 5 3 2}

In the quantum field theory side, there are many different but equivalent
bases for the quarter and eighth BPS operators. These bases can be
transformed to each. A way to describe excitations is to use their
collective variables, and this includes coherent states of collective
excitations. One of the methods is by coherent state basis, written in terms
of displacer, $\mathrm{exp}(\sum\limits_{n,m}N_{n,m}^{-1}\Lambda
_{n,m}a_{n,m}^{\dagger })\left\vert \Omega \right\rangle $. Here $\left\vert
\Omega \right\rangle $ is a background state, which can also be a vacuum
state e.g. $\left\vert 0\right\rangle \ $as a special case. The $\Lambda
_{n,m}$ is the coherent state amplitude and the eigenvalue of $a_{n,m}$ on
the coherent states. The $a_{n,m}^{\dagger }$ adds a column of length $n$ to
YT$_{1}$ and a column of length $m$ to YT$_{2}$. These can be viewed as the
generalization of the coherent states in \cite{Balasubramanian:2018yjq}. By
the Berry curvature formalism, the symplectic form is equivalent to $\Omega
\sim \sum\limits_{n,m}\delta \Lambda _{n,m}\wedge \delta \bar{\Lambda}_{n,m}$%
. This describes the symplectic form in the configuration space of states
labelled by coherent state basis.

The symplectic forms can be written in other operator bases. The symplectic
forms constructed by different bases, including operator bases and geometric
bases \cite{Grant:2005qc,Maoz:2005nk,Balasubramanian:2018yjq}, can be
transformed to each other by changes of variables and canonical
transformations.

To describe the superstar microstates, it is natural to use states labeled
by the density function, as exposited in \cite{Balasubramanian:2005mg} and
more recently in \cite{Balasubramanian:2018yjq} in relation to eigenstate
thermalization hypothesis, among many other important aspects. We consider a
coarse density $u_{0}$. Further, we consider deformations around this
distribution, which are $\delta u_{0}$, and we analyzed them in Sec. \ref%
{sec 5 3 1}. The deformed density is $u_{0}+\delta u_{0}$, with general
variations.

The change of density, is related to the dilation and squeezing of white
regions. The dilation of white regions decreases the density of the black
regions, and vice versa, the squeezing of white regions increases the
density of the black regions, as they are complementary regions as explained
in Sec. 2. Now, dilation of white regions is related to adding YT columns.

Denote $c_{1,i},c_{2,i}\ $as the numbers of columns with length $i$, for the
two YTs respectively. $a_{n,m}^{\dagger }$ adds a column of length $n$ to YT$%
_{1}$ and a column of length $m$ to YT$_{2}$. These operations make shape
deformations of the two YTs. Let\ $x$ label rows, $y_{1},y_{2}$ label
columns. The ideas have been introduced in \cite%
{Balasubramanian:2005mg,Balasubramanian:2018yjq}. The typical tableaux are
described by the shape functions $y_{1}(x),y_{2}(x)$, and the slopes are the
derivatives,
\begin{equation}
u_{0}\simeq \frac{1}{1+y_{1}^{\prime }+y_{2}^{\prime }}.
\end{equation}%
This is associated to (\ref{eqn 55 02}). Consider the deformation from $%
u_{0}\ $to $u_{0}+\delta u(r^{2})$. The change of the shape functions are $%
\delta y_{1}(x),\delta y_{2}(x)$, and $y_{1}(x_{1})+x_{1}+y_{2}(x_{2})+x_{2}=%
\frac{1}{2}r^{2}$. By variation principle we have%
\begin{equation}
\delta u(r^{2})\simeq -\frac{1}{(1+y_{1}^{\prime }+y_{2}^{\prime })^{2}}%
\left( \delta y_{1}^{\prime }(x_{0,1})+\delta y_{2}^{\prime
}(x_{0,2})\right) .
\end{equation}%
Now we make an approximation that the YTs are approximately triangular, so
that $y_{1}(x),y_{2}(x)$ are approximately linear in $x$, and hence we
neglect the$\ y_{1}^{\prime \prime }(x),y_{2}^{\prime \prime }(x)$ terms
which are approximately zero.

Observe that $y_{1}(x)\simeq \omega _{1}x$ where $\omega _{1}\simeq \frac{%
N_{c,1}}{N}$, and $y_{2}(x)\simeq \omega _{2}x$ where $\omega _{2}\simeq
\frac{N_{c,2}}{N}$. Since $\delta y^{\prime }\left( x_{0}\right) $ means
changing the number of columns at $x_{0}$, and since columns are viewed as
holes, reducing the holes increases particle density, and vice versa. Hence $%
\delta y_{1}^{\prime }(x_{0,1})+\delta y_{2}^{\prime }(x_{0,2})=\delta
\langle c_{1,N-x_{0,1}}\rangle +\delta \langle c_{2,N-x_{0,2}}\rangle $.
Denote $\omega =\omega _{1}+\omega _{2}$,
\begin{equation}
\delta u(r^{2})\simeq -\frac{1}{(1+\omega )^{2}}\left( \delta \langle
c_{1,N-x_{0,1}}\rangle +\delta \langle c_{2,N-x_{0,2}}\rangle \right) .
\end{equation}%
Hence$\ \delta u_{n,m}\simeq -N_{n,m}^{-1}(\delta \langle
c_{1,N-x_{0,1}}\rangle +\delta \langle c_{2,N-x_{0,2}}\rangle )$.$\ $For the
operators with appropriate normalization of two-point functions, we have
that $\alpha _{n,m}$ is eigenvalue of$\ a_{n,m}$. We have that$\
c_{1,N-x_{0}}+c_{2,N-x_{0}}\sim \langle a_{n,m}^{\dagger }a_{n,m}\rangle $.
Writing by using variable $\alpha _{n,m}=|\alpha _{n,m}|e^{i\pi _{n,m}}$, we
have that the canonical conjugate variable to $u_{n,m}$ is the canonical $%
\pi _{n,m}$. Hence by canonical transformations, we have that%
\begin{equation}
\Omega \sim \int \delta u(r)\wedge \delta \pi (r)\sim \sum_{n,m}\delta
u_{n,m}\wedge \delta \pi _{n,m}\sim \sum_{n,m}\delta \Lambda _{n,m}\wedge
\delta \bar{\Lambda}_{n,m}.
\end{equation}%
That is, the symplectic form written using density functions and the
symplectic form written using coherent state amplitudes, are equivalent to
each other, by canonical transformations.

These different writings of symplectic forms are related to each other by
canonical transformations, and changes of variables, including Fourier
transforms. These modes labeled by $n,m$ can also be captured by computing
correlators. In Appendix E, we also discuss the symplectic forms in special
cases. It would be good to understand more details of the canonical
transformations between symplectic forms obtained by operator bases and
geometric bases. We leave more detailed calculations and will revisit these
subjects in future investigations.

\section{Discussion}

\label{sec_discussion}

We obtained generalized Laplace-type equations with sources, from linearized
Monge-Ampere equations, on a general reference geometry and for the
asymptotically AdS geometry. This led us to obtain asymptotically AdS
solutions. One may have further simplifications for above cases. They can be
used in combination with a stretched limit, where boundary conditions can be
placed at the stretched surface. These boundary conditions include grey
droplets.

The above stretched limit, e.g. (\ref{eqn 66}), is naturally used for the
superstar, which has a natural cut-off surface \cite{Balasubramanian:2018yjq}%
. We place grey droplet boundary conditions on the stretched surface. The
two-charge superstar is the extremal case of the near-extremal black hole.
The superstar corresponds to grey droplet boundary conditions on the
stretched surface. On the gravity side, geometric quantization and
coarse-graining can be naturally combined, and this has also been performed
in \cite{Balasubramanian:2018yjq}. The coarse gaining of the density
function, leads to the coarse-graining of symplectic forms.

For coarse-graining purpose, we primarily use the momentum space version and
the associated Fourier transforms. One may also use Mellin transforms to
analyze the solutions. We can alternatively solve the global solution by
Mellin transforms. We can also perform the analysis using the methods of
\cite{Skenderis:2007yb,Christodoulou:2016nej}. Among many other aspects, the
solution techniques and their various relations have been summarized in Sec. %
\ref{sec 2 3}, Sec. \ref{sec 5 2}, and Appendix \ref{sec F}.

The IR part of these geometries \cite{Lin:2004nb} as well as smooth quarter
BPS geometries, with a cutoff away from the asymptotic boundary, are also
dual to the IR of the irrelevant deformation of SCFTs. The IR region of the
geometry, in particular, is for the IR region of the geometry dual to the
irrelevant deformation preserving quarter BPS supersymmetry, and half BPS
supersymmetry, respectively \cite%
{Cordova:2016emh,Caetano:2020ofu,Cordova:2016xhm}. See also related aspects
\cite{Roychowdhury:2023hvq}. These deformations are deformations by BPS
irrelevant operators. It would be good to understand these aspects in more
details.

There are many hierarchies of scales in smooth bubbling geometries, partly
due to large ratios of different sizes of different coexisting droplets or
bubbles, and partly due to that the spectra of the QFT side have rich
phenomena as well as complexity; and there can be many intermediate scales
in between the fundamental string scale and AdS scale. The coarse-graining
scale is a scale that separates the high energy with the low energy. The
black hole microstructure can appear and be evident at scales that are much
larger than the planck scale, e.g. \cite%
{Balasubramanian:2018yjq,Nomura:2019qps}.

The systems can be put inside top-down string theory and are UV complete.
The string theory describes hard modes very sophisticatedly, because of its
UV completeness. On the other hand, there are no UV or IR divergences in
full string theory, e.g. \cite{Witten:2012bh}. Soft modes in the bulk are
included in the description in the dual field theory side, e.g. \cite%
{Balasubramanian:2018yjq,Nomura:2019qps}. See also related discussion on
soft modes related to various stretched horizons \cite{Nomura:2019qps}. Our
work has also implications for soft modes on the stretched horizons, as
discussed in e.g. Appendix E.

Infalling massless particles can be trapped into the interior regions in
droplet space without escaping \cite{Berenstein:2023vtd}, which is similar
to entering into a maze. The behavior is similar to behavior in fuzzballs.
These scenarios as well as our related discussions, are closely related to
aspects of microstate geometries and microstructure of fuzzballs and
blackholes, e.g. \cite{Mayerson:2020tpn}-\cite{Bena:2022ldq}. Various
insights and effects of microstate geometries have been unraveled. See also
related discussions \cite{deMelloKoch:2020jmf,McLoughlin:2020zew} on chaotic
behaviors.

The superstar \cite{Myers:2001aq} is the extremal case of the near-extremal
black hole. There is a near-extremality parameter which is related to the
near-BPS energy. These are related to near-BPS states, e.g. \cite%
{Balasubramanian:2007bs,Fareghbal:2008ar}. One can obtain non-BPS
excitations as further excitations on top of BPS excitations. Near BPS
states can also describe near BPS black holes \cite%
{Behrndt:1998jd,Behrndt:1998ns,Balasubramanian:2007bs,Fareghbal:2008ar}, and
are related to various giant configurations. The string configurations \cite%
{Berenstein:2022cju,Holguin:2021qes} on giants are relevant. One of the very
useful methods for near-BPS states, is the spin-matrix theory, see \cite%
{Baiguera:2021hky}-\cite{Harmark:2016cjq} and related references. See also
related aspects and methods \cite{Lin:2022wdr}. Another one of very useful
methods for near-BPS states, are restricted Schurs of giant graviton branes,
e.g. \cite{deMelloKoch:2012ck,Carlson:2011hy} and related references. They
also include BPS states in the moduli space.

String probes in these backgrounds can be conveniently computed as they are
globally smooth. Excitations, for example string excitations as well as
light field excitations on the bubbling geometries, have been analyzed in
e.g. \cite{Aguilera-Damia:2017znn,Chen:2007gh,deMelloKoch:2018tlb}, and
their related references.

\section*{Acknowledgments}

We would like to thank V. Balasubramanian, A. Belin, I. Bena, D. Berenstein,
J. Caetano, P. Caputa, Y.J. He, Z. He, J. Hou, L.Y. Hung, A. Jevicki, R. de
Mello Koch, M.K. Kim, S. Komatsu, Y. Jiang, Y.Z. Li, J. Liu, H. Lu, O.
Lunin, J. Maldacena, J. Nian, C. Nunez, T. Okazaki, T. Okuda, K.
Papadodimas, M. Paulos, S. Ramgoolam, L. Rastelli, V.S. Rychkov, K.
Skenderis, J. Simon, M. Taylor, N. Warner, for related discussions in the
past, and R. Suzuki for crucial comments and correspondences. The work was
supported in part by National Key R\&D Program of China grant
2020YFA0713000, by Grant No. TH-533310008, by Overseas high-level talents
program, by Fundamental Research Funds for the Central Universities of
China, by Grant No. 4007012320, and by Grant No. 3207012204.

\appendix

\section{AdS background solution and expansion}

\label{sec A} \renewcommand{\theequation}{A.\arabic{equation}} %
\setcounter{equation}{0} \renewcommand{\thethm}{A.\arabic{thm}} %
\setcounter{thm}{0} \renewcommand{\theprop}{A.\arabic{prop}} %
\setcounter{prop}{0}

Here we explain how to use the $AdS_{5}\times S^{5}$ as a reference
geometry, as an example, for the purpose of performing linearized technique,
among other things.

The potential with the AdS radius $l$ restored is ($l$ was sometimes set as
1 in \cite{Chen:2007du} for simplicity and it can be restored by dimensional
analysis or dimensional counting)
\begin{eqnarray}
K_{(0)} &=&\frac{1}{2}\left( \frac{1}{2}(r^{2}+y^{2}+l^{2})+\sqrt{\frac{1}{4}%
(r^{2}+y^{2}-l^{2})^{2}+y^{2}l^{2}}\right)  \notag \\
&&-\frac{1}{2}l^{2}\log \left( \frac{1}{2}(r^{2}+y^{2}+l^{2})+\sqrt{\frac{1}{%
4}(r^{2}+y^{2}-l^{2})^{2}+y^{2}l^{2}}\right)  \notag \\
&&-\frac{1}{2}y^{2}\log \left( \frac{1}{2}(-r^{2}+y^{2}+l^{2})+\sqrt{\frac{1%
}{4}(r^{2}+y^{2}-l^{2})^{2}+y^{2}l^{2}}\right) +\frac{1}{2}y^{2}\log (y).
\notag \\
&&
\end{eqnarray}%
Now we consider $r^{2}+y^{2}>l^{2}$ (this is outside and away from the
central droplet) and expand the potential in powers of $l^{2}$. We have
\begin{equation}
\sqrt{\frac{1}{4}(r^{2}+y^{2}-l^{2})^{2}+y^{2}l^{2}}=\frac{1}{2}%
(r^{2}+y^{2}-l^{2})+\frac{y^{2}}{(r^{2}+y^{2})}l^{2}+O(l^{4}).
\end{equation}%
Hence
\begin{eqnarray}
K_{(0)} &=&\frac{1}{2}\left( r^{2}+y^{2}+\frac{y^{2}}{(r^{2}+y^{2})}%
l^{2}\right)  \notag \\
&&-\frac{1}{2}l^{2}\log \left( r^{2}+y^{2}+\frac{y^{2}}{(r^{2}+y^{2})}%
l^{2}\right)  \notag \\
&&-\frac{1}{2}y^{2}\log \left( y^{2}+\frac{y^{2}}{(r^{2}+y^{2})}l^{2}\right)
+\frac{1}{2}y^{2}\log (y)+O(l^{4}).
\end{eqnarray}%
The $o(l^{2})$ term $\frac{y^{2}}{2(r^{2}+y^{2})}l^{2}$ from the first term
and the $o(l^{2})$ term $-\frac{y^{2}}{2(r^{2}+y^{2})}l^{2}$ from the second
log term precisely cancel, and we thus have
\begin{eqnarray}
K_{(0)} &=&\frac{1}{2}\left( r^{2}+y^{2}\right) -\frac{1}{2}l^{2}\log \left(
r^{2}+y^{2}\right) -\frac{1}{2}y^{2}\log \left( y^{2}\right) +\frac{1}{2}%
y^{2}\log (y)+O(l^{4})  \notag \\
&=&\frac{1}{2}\left( r^{2}+y^{2}\right) -\frac{1}{2}l^{2}\log \left(
r^{2}+y^{2}\right) -\frac{1}{4}y^{2}\log (y^{2})+O(l^{4}).  \label{eqn 05_}
\end{eqnarray}%
The first three terms are most important terms and are leading asymptotic
behaviors.

\section{Multi droplets, dilute droplets, and the dilute distribution regime}

\label{sec B} \renewcommand{\theequation}{B.\arabic{equation}} %
\setcounter{equation}{0} \renewcommand{\thethm}{B.\arabic{thm}} %
\setcounter{thm}{0} \renewcommand{\theprop}{B.\arabic{prop}} %
\setcounter{prop}{0}

We consider isolated small smooth droplets, which are very far from each
other. Near the isolated small round droplet, the geometry locally
approaches that of smaller AdS space, with $l^{2}$ parameter effectively
rescaled to $h_{i}^{2}$. The small droplet is dual to giants, near each
small isolated droplet.

We have
\begin{eqnarray}
Z &=&\frac{|z_{1}|^{2}+|z_{2}|^{2}+y^{2}-L_{0}^{2}}{2\sqrt{%
(|z_{1}|^{2}+|z_{2}|^{2}+y^{2}-L_{0}^{2})^{2}+4y^{2}L_{0}^{2}}}  \notag \\
&&+\sum\limits_{i=2}^{n}\left( \frac{%
|z_{1}-s_{1i}|^{2}+|z_{2}-s_{2i}|^{2}+y^{2}-h_{i}^{2}}{2\sqrt{%
(|z_{1}-s_{1i}|^{2}+|z_{2}-s_{2i}|^{2}+y^{2}-h_{i}^{2})^{2}+4y^{2}h_{i}^{2}}}%
-\frac{1}{2}\right) ,  \label{eqn 29}
\end{eqnarray}%
in the dilute distribution regime. The extreme-ratio regime is $h_{i}^{2}\ll
L^{2}$, for $i=2,...,n$. And $h_{1}^{2}=L_{0}^{2}.$ The droplets are
well-separated. We also include the central droplet, whose center is denoted
as $s_{11}=s_{21}=0$. We can also write this expression for $Z-\frac{1}{2}$
which has a superimposition structure. The $V_{(1)}=Z_{(1)}=Z-Z_{0}$ also
has a superimposition structure. Locally near the small droplet, the local
behavior is similar to the local behavior of the central droplet. Locally,
the small droplet is dual to$~n_{i}=h_{i}^{2}N/L^{2}$ three-brane giants.
The quanta for each small droplet $q_{i}$ is much smaller than the total
quanta $N$, i.e. $q_{i}\ll N$; and this is referred to as the extreme-ratio
limit. Dilute means that the droplets are isolated from each other.

Now we expand in the regime $r^{2}+y^{2}\gg h_{i}^{2}$. The Kahler potential
is
\begin{eqnarray}
K &\approx &\sum\limits_{i=1}^{n}\frac{1}{2}\left(
|z_{1}-s_{1i}|^{2}+|z_{2}-s_{2i}|^{2}+y^{2}\right)  \notag \\
&&-\sum\limits_{i=1}^{n}\frac{1}{2}h_{i}^{2}\log \left(
|z_{1}-s_{1i}|^{2}+|z_{2}-s_{2i}|^{2}+y^{2}\right) -\frac{1}{4}y^{2}\log
(y^{2})+c.
\end{eqnarray}%
where $s_{11}=s_{21}=0$. Hence we have
\begin{eqnarray}
K &\approx &\sum\limits_{i=1}^{n}\frac{1}{2}\left(
|z_{1}|^{2}+|z_{2}|^{2}+y^{2}\right)  \notag \\
&&-\sum\limits_{i=1}^{n}\frac{1}{2}h_{i}^{2}\log \left(
|z_{1}-s_{1i}|^{2}+|z_{2}-s_{2i}|^{2}+y^{2}\right)  \notag \\
&&-\frac{1}{4}y^{2}\log (y^{2})+c.
\end{eqnarray}%
$h_{i}^{2}$ is proportional to the total number of eigenvalues $n_{i}$ in
the droplet $i$. The droplet is an eigenvalue cluster.

With appropriate rescaling,
\begin{eqnarray}
&&K|_{y=0}\sim \frac{1}{2}\left( |z_{1}|^{2}+|z_{2}|^{2}\right)
-\sum\limits_{i=1}^{n}\frac{1}{2}n_{i}\log \left(
|z_{1}-s_{1i}|^{2}+|z_{2}-s_{2i}|^{2}\right) +c.  \notag \\
&&  \label{eqn}
\end{eqnarray}%
This behaves as a potential that a probe eigenvalue labeled by $z_{i}$,$\ $%
feels in the presence of other eigenvalues, in which there is a logarithmic
potential. We have $\sum\limits_{i=1}^{n}n_{i}=N$. The relation between $%
n_{i},h_{i}$ is $n_{i}\sim h_{i}^{2}$. There is also a confining potential $%
|z_{i}|^{2}$, and the confining potential in the dual field theory side has
been originally derived from dual field theory dynamics \cite%
{Berenstein:2005aa}. The droplet configuration has multiple equivalent
interpretations from the dual field theory side. See also e.g. \cite%
{Berenstein:2007wz,deMelloKoch:2016whh}. The droplet configurations in the
gravity side correspond to eigenvalue clusters in the field theory side.%
\footnote{%
See also various related discussions e.g. \cite{Berenstein:2008eg}-\cite%
{Berenstein:2008jn} and their related aspects in diverse or different
contexts. There are rich dynamics, phases, and strong-coupling phenomena
that are inter-connected.}

This is in the dilute distribution limit, and the extra isolated
eigenvalues, are well-separated. There can be more general regimes than the
regime we consider here. The potential on the droplet space, e.g. (\ref{eqn}%
) is related to the logarithm of the wavefunction inner-product of
appropriate matrix wavefunction basis of the field theory side. On the other
hand, matrix wavefunction bases may not be unique, and there may also be
other bases by changes of variables.

We further confirm the observations in \cite%
{Berenstein:2005aa,Berenstein:2007wz} and their related aspects. Our case
here is a special case of \cite{Berenstein:2005aa,Berenstein:2007wz}.

\section{Superstar and asymptotic expansion}

\label{sec C} \renewcommand{\theequation}{C.\arabic{equation}} %
\setcounter{equation}{0} \renewcommand{\thethm}{C.\arabic{thm}} %
\setcounter{thm}{0} \renewcommand{\theprop}{C.\arabic{prop}} %
\setcounter{prop}{0}

The two-charge superstar have also been analyzed in \cite{Chen:2007du}. The
symbols $r,{\tilde{r}}$ in Sec. 6.2.1 of \cite{Chen:2007du} are denoted as ${%
\tilde{r},}r$ here, due to that we use symbol $r$ for the droplet space. The
$Z$ function for superstar is
\begin{eqnarray}
Z &=&\frac{1}{2}-\frac{1}{1+Q_{2}+Q_{3}+{\tilde{r}}^{2}+{Q_{2}\,Q_{3}}/{%
\tilde{r}^{2}}+\cot ^{2}\theta \lbrack ({\tilde{r}}^{2}+Q_{3})\sin
^{2}\alpha +({\tilde{r}}^{2}+Q_{2})\cos ^{2}\alpha ]}\ .  \notag \\
&&
\end{eqnarray}%
We work in unit $l=1$. Note the \textquotedblleft $1$\textquotedblright\ on
the right hand side correspond to $l^{2}$, when restored.

The analysis is that we expand from the point of view of asymptotic region.
Consider the identification for large ${\tilde{r}}$ and
\begin{equation}
y={\tilde{r}}\sin \theta ,\ r={\tilde{r}}\cos \theta .
\end{equation}%
We have $r^{2}={\tilde{r}}^{2}-y^{2}$ and$\ \sin ^{2}\theta =\frac{y^{2}}{{%
\tilde{r}}^{2}}$.$\ $Consider equal charge case for simplicity, $%
Q_{2}=Q_{3}=Q$. We have
\begin{eqnarray}
Z &=&\frac{1}{2}-\frac{\sin ^{2}\theta }{\sin ^{2}\theta \left( 1+2Q+{\tilde{%
r}}^{2}+{Q}^{{2}}/{\tilde{r}}^{2}\right) +\cos ^{2}\theta ({\tilde{r}}^{2}+Q)%
}  \notag \\
&=&\frac{1}{2}-\frac{\sin ^{2}\theta }{{\tilde{r}}^{2}+Q+\sin ^{2}\theta
(1+Q)+\sin ^{2}\theta {Q}^{{2}}/{\tilde{r}}^{2}}.
\end{eqnarray}%
We use $1/{\tilde{r}}$ expansion. The large ${\tilde{r}}$ expansion is%
\begin{equation}
Z\simeq \frac{1}{2}-\frac{\sin ^{2}\theta }{{\tilde{r}}^{2}}+\frac{\sin
^{2}\theta (Q+\sin ^{2}\theta (1+Q))}{{\tilde{r}}^{4}}+...
\end{equation}%
and%
\begin{equation}
\frac{Q+\sin ^{2}\theta (1+Q)}{{\tilde{r}}^{2}}=\frac{Q}{{\tilde{r}}^{2}}%
+(1+Q)\frac{y^{2}}{{\tilde{r}}^{4}}.
\end{equation}%
To extract the distribution, we have then,%
\begin{equation}
Z-\frac{1}{2}\simeq -\frac{y^{2}l^{2}}{(r^{2}+y^{2})^{2}}\left( 1-\frac{Q}{%
r^{2}+y^{2}}\right) +o(\frac{1}{(r^{2}+y^{2})^{4}}),  \label{eqn 63}
\end{equation}%
with asymptotic AdS radius $l$ restored. The superstar metric and smooth
geometries differ slightly in the interior region approaching the stretched
horizon. We expand in the exterior region.

We use a technique \cite{Kimura:2011df} that we expand using $%
1/(r^{2}+y^{2}) $.$\ $We can expand using $r^{2}+y^{2},$ by the dilute
distribution configuration, e.g. (\ref{eqn 29}),
\begin{eqnarray}
&&\frac{r^{2}+y^{2}-h_{i}^{2}}{2\sqrt{%
(r^{2}+y^{2}-h_{i}^{2})^{2}+4y^{2}h_{i}^{2}}}-\frac{1}{2}  \notag \\
&=&-\frac{y^{2}h_{i}^{2}}{(r^{2}+y^{2})^{2}}+o(h_{i}^{4}).
\end{eqnarray}%
It has both $r^{2}+y^{2}$ dependence and $y^{2}/r^{2}$ angle dependence. We
have%
\begin{eqnarray}
&&-\frac{y^{2}h_{i}^{2}}{(|z_{1}-s_{1i}|^{2}+|z_{2}-s_{2i}|^{2}+y^{2})^{2}}
\notag \\
&\simeq &-\frac{y^{2}h_{i}^{2}}{(r^{2}+y^{2})^{2}}\left( 1-\frac{%
2(|s_{1i}|^{2}+|s_{2i}|^{2})}{r^{2}+y^{2}}\right) +o(\frac{1}{%
(r^{2}+y^{2})^{4}})  \notag \\
&\simeq &-\frac{y^{2}h_{i}^{2}}{{\tilde{r}}^{4}}\left( 1-\frac{%
2(|s_{1i}|^{2}+|s_{2i}|^{2})}{{\tilde{r}}^{2}}\right) +o(\frac{1}{{\tilde{r}}%
^{8}}).  \label{eqn 64}
\end{eqnarray}%
Hence by (\ref{eqn 63}) and (\ref{eqn 64}),
\begin{eqnarray}
&&-\frac{y^{2}l^{2}}{{\tilde{r}}^{4}}\left( 1-\frac{Q}{{\tilde{r}}^{2}}%
\right)  \notag \\
&\simeq &\sum\limits_{i}-\frac{y^{2}h_{i}^{2}}{{\tilde{r}}^{4}}\left( 1-%
\frac{2(|s_{1i}|^{2}+|s_{2i}|^{2})}{{\tilde{r}}^{2}}\right) .  \label{eqn 65}
\end{eqnarray}%
We include central droplet $i=0$, and $h_{0}^{2}=l_{0}^{2}$.\ Hence by (\ref%
{eqn 65}), $\sum\limits_{i}h_{i}^{2}=l^{2}$. Hence%
\begin{equation}
l^{2}Q\simeq \sum\limits_{i}2h_{i}^{2}(|s_{1i}|^{2}+|s_{2i}|^{2}).
\end{equation}%
Note this is in the dilute distribution configuration limit. The small
droplets are far away from the central droplet, $|s_{1i}|^{2}+|s_{2i}|^{2}%
\gg 1.$

We assumed equal charge, and when we restore $Q_{2}$ and $Q_{3}$, we find,
\begin{equation}
\frac{1}{2}l^{2}Q_{2}\simeq \sum\limits_{i}2h_{i}^{2}|s_{1i}|^{2},\ \ \ \
\frac{1}{2}l^{2}Q_{3}\simeq \sum\limits_{i}2h_{i}^{2}|s_{2i}|^{2}.
\label{eqn 67}
\end{equation}%
Note the convention of labels $2,3$ in \textquotedblleft $Q_{2,3}$%
\textquotedblright $\ $in \cite{Chen:2007du} are labels $1,2$ here in this
paper. One can also consider equal size cases where $h_{i}^{2}=h^{2}$ are
the same for each droplets, and take a continuous distribution limit, under
which the droplet density becomes grey distributions.

The geometries are asymptotic to $AdS_{5}\times S^{5}$. We set the $AdS$
radius parameter to be $l$ for this $AdS_{5}\times S^{5}$ geometry at the
asymptotia.

\section{Properties of $F_{+}(k,y)$ and a new approximate solution}

\renewcommand{\theequation}{D.\arabic{equation}} \setcounter{equation}{0} %
\renewcommand{\thethm}{D.\arabic{thm}} \setcounter{thm}{0} %
\renewcommand{\theprop}{D.\arabic{prop}} \setcounter{prop}{0}

Denote the solution for \textquotedblleft $y=0$ boundary\textquotedblright\
case as $F_{+}(k,y)$. We can have approximate solution as follows:
\begin{equation}
F_{+}(k,y)\simeq c_{1}\frac{1}{2}k^{2}y^{2}K_{2}(ky)+\frac{c_{2}l^{2}}{%
y_{c}^{2}+y^{2}}\frac{1}{48}k^{4}y^{4}K_{4}(ky),\ \ \ y\geq 0.
\label{eqn 58}
\end{equation}%
Note the above \textquotedblleft $y=y_{c}$ boundary\textquotedblright\
solution (\ref{eqn 54}) is:
\begin{equation}
F(k,y)\simeq c_{1}\frac{1}{2}k^{2}y^{2}K_{2}(ky)+c_{2}\frac{1}{48}%
k^{4}l^{2}y^{2}K_{4}(ky).  \label{eqn 57}
\end{equation}%
When $y\rightarrow 0$:
\begin{equation}
F_{+}(k,y)\simeq c_{1}\frac{1}{2}k^{2}y^{2}K_{2}(ky)+\frac{c_{2}l^{2}}{%
y_{c}^{2}}\frac{1}{48}k^{4}y^{4}K_{4}(ky),\ \ \ 10^{-1}y_{c}>y\geq 0.
\end{equation}%
$F_{+}(k,y)|_{y=0}=c_{1}+\frac{c_{2}l^{2}}{y_{c}^{2}}=1$. This means it
satisfies the boundary condition placed at $y=0$, approaching delta function.

When $y>10y_{c}$:%
\begin{equation}
F_{+}(k,y)\simeq c_{1}\frac{1}{2}k^{2}y^{2}K_{2}(ky)+c_{2}\frac{1}{48}%
k^{4}l^{2}y^{2}K_{4}(ky),\ \ \ y>10y_{c}.
\end{equation}%
This becomes (\ref{eqn 57}). Hence (\ref{eqn 58}) is an approximate solution
for \textquotedblleft $y=0$ boundary\textquotedblright\ case, satisfying the
boundary condition placed at $y=0$.

\section{Relation to the symplectic form and discussion}

\renewcommand{\theequation}{E.\arabic{equation}} \setcounter{equation}{0} %
\renewcommand{\thethm}{E.\arabic{thm}} \setcounter{thm}{0} %
\renewcommand{\theprop}{E.\arabic{prop}} \setcounter{prop}{0}

We have observed similar phenomena encountered in \cite%
{Balasubramanian:2018yjq} in the context of deformation modes in smooth
bubbling geometries and their behaviors under coarse-graining. We first work
with smooth droplet configurations, and later perform coarse-graining.

The coordinate $z_{1}=q_{1}+ip_{1},z_{2}=q_{2}+ip_{2}$, can be identified
with the holomorphic coordinate associated to the degree of freedom for a
probe eigenvalue in the field theory side. For related discussions and
equivalent interpretations, see e.g. \cite{Berenstein:2022srd}. This can be
identified as one particle phase space. Hence one can identify the 4d
droplet space coordinate as one particle phase space. We can also write $%
z_{1}=r_{1}e^{i\phi },z_{2}=r_{2}e^{i\varphi },r^{2}=|z_{1}|^{2}+|z_{2}|^{2}$%
. We can denote $|z_{1}|=r\cos \chi $, $|z_{2}|=r\sin \chi $, and $0\leq
\chi \leq \frac{\pi }{2}$, $\sin \chi \geq 0,\cos \chi \geq 0$.

Here we mainly use coherent states and Berry curvature formalism. One can
also use coadjoint orbit formalism, coherent state formalism, and other
formalisms \cite{Puta:}. It may also be interesting to use collective field
theory. The calculations can be simplified and unified by the Berry
curvature formalism.

As discussed in e.g. \cite{Balasubramanian:2018yjq} and Sec. \ref{sec 5 3 2}
of this paper, $u(z)$ is density, and we define $\pi (z)$ as the canonical
conjugate momentum to the density $u(z)$. The symplectic form is%
\begin{equation}
\boldsymbol{\Omega }=\int_{r<R}d^{4}z\,\boldsymbol{\delta }\pi (z)%
\boldsymbol{\delta }u(z).
\end{equation}%
We use canonical quantization. There is a natural Poisson structure induced
by the two-dimensional quantum harmonic oscillator \cite{Puta:}.

Consider the action of coherent state operators on excitations of a
reference state denoted by $|\psi _{0}\rangle $. The resulting state can be
expressed as
\begin{equation}
|\psi \rangle =e^{iT_{\pi }}|\psi _{0}\rangle ,\;\;\;T_{\pi }=\int
\prod_{i=1,2}\frac{dp_{i}^{\prime }dq_{i}^{\prime }}{2\pi \hbar }\pi
(q_{i}^{\prime },p_{i}^{\prime })u(q_{i}^{\prime },p_{i}^{\prime }).
\end{equation}%
The change of the state is given by
\begin{equation}
\delta |\psi \rangle =i\int \prod_{i=1,2}\frac{dp_{i}^{\prime
}dq_{i}^{\prime }}{2\pi \hbar }\delta \pi (q_{i}^{\prime },p_{i}^{\prime
})u(q_{i}^{\prime },p_{i}^{\prime })|\psi _{0}\rangle .
\end{equation}%
This characterizes the coadjoint orbit of $\psi _{0}$, which gives the orbit
of the coadjoint action. The symplectic form above can also be expressed as $%
\hbar $ times the Berry curvature. This has also been discussed in the
context of gauge/gravity correspondence by \cite{Belin:2018fxe} in general.
We have, see also \cite{Belin:2018fxe} and \cite{Balasubramanian:2018yjq},
hence that
\begin{equation}
\Omega =\hbar F_{Berry}=i\hbar \left( \langle \delta _{1}\psi |\delta
_{2}\psi \rangle -\langle \delta _{2}\psi |\delta _{1}\psi \rangle \right) .
\end{equation}%
Hence,
\begin{equation}
\Omega =\int \prod_{i=1,2}\frac{dp_{i}^{\prime }dq_{i}^{\prime }}{2\pi \hbar
}\delta \pi (q_{i}^{\prime },p_{i}^{\prime })\delta u(q_{i}^{\prime
},p_{i}^{\prime }).
\end{equation}

We work in the natural units. Our methods here are similar to those of \cite%
{Balasubramanian:2018yjq}. We consider concentric shell model, e.g. \cite%
{Kimura:2011df}; we consider the black regions are along the radial
directions in concentric shell configurations. The shells are labeled by $i$%
. The $u$ depend on the droplet pattern. We have%
\begin{equation}
u(r)=\sum_{i}(-1)^{i}\Theta (r_{i}-r),
\end{equation}%
where $\Theta (x)$ denotes step function. The $i$ is $i$-th droplet shell.
For $i$ even, the black droplet region is between $r_{i}$ and $r_{i-1}$. The
droplet spacing along $r$ is denoted as $\epsilon .\ $The change of density
localize on droplet edge $i.$ This is a choice of basis for deformation
modes. Consider the change of the density from $u$ to $u+\delta u$, and%
\begin{equation}
\delta u(r,\phi ,\varphi )=\sum_{i}(-1)^{i}\delta r_{i}(\phi ,\varphi
)\delta (r_{i}-r).
\end{equation}%
We have $u_{i}(\phi ,\varphi )$ and
\begin{equation}
\delta u(r,\phi ,\varphi )=\epsilon \sum_{i}\delta u_{i}(\phi ,\varphi
)\delta (r-r_{i}).
\end{equation}%
The change of density $\boldsymbol{\delta }u(r,\phi ,\varphi )$ is
decomposed into $\delta u_{i}(\phi ,\varphi )$. We than have
\begin{equation}
\delta r_{i}(r,\phi ,\varphi )=(-1)^{i}\epsilon \delta u_{i}(r,\phi ,\varphi
).
\end{equation}%
We then get $\partial _{\phi }\delta \pi (r_{i},\phi ,\varphi )=\cos
^{2}\chi r_{i}\delta r_{i}$, $\partial _{\varphi }\delta \pi (r_{i},\phi
,\varphi )=\sin ^{2}\chi r_{i}\delta r_{i}$, and hence
\begin{equation}
\delta \pi (r,\phi ,\varphi )\sim \frac{1}{4\pi }\int d\phi ^{\prime
}d\varphi ^{\prime }(\mathrm{Sign}(\phi -\phi ^{\prime })\cos ^{2}\chi +%
\mathrm{Sign}(\varphi -\varphi ^{\prime })\sin ^{2}\chi )\sum_{i}\epsilon
r_{i}(-1)^{i}\delta u(r_{i},\phi ^{\prime },\varphi ^{\prime })H_{\epsilon
}(r-r_{i}).
\end{equation}%
Here $H_{\epsilon }(r-r_{i})$ is a smooth function covering the $i$-th
shell; it's centered around $r_{i}$ with width $\epsilon $, and its value is
1 in between $r_{i}-\frac{1}{2}\epsilon $ and $r_{i}+\frac{1}{2}\epsilon $,
and drops to zero outside this range. We have $r_{i}=i\epsilon $, and hence $%
e^{i\pi r_{i}/\epsilon }H_{\epsilon }(r-r_{i})=(-1)^{i}H_{\epsilon
}(r-r_{i}) $. Hence,
\begin{eqnarray}
\Omega &\sim &-\frac{1}{16\pi ^{3}}\int r^{2}dr\int \sin \chi \cos \chi
d\chi d\phi d\varphi d\phi ^{\prime }d\varphi ^{\prime }\boldsymbol{\delta }%
u(r,\phi ,\varphi )(\mathrm{Sign}(\phi -\phi ^{\prime })\cos ^{2}\chi +%
\mathrm{Sign}(\varphi -\varphi ^{\prime })\sin ^{2}\chi )  \notag \\
&&\sum_{i}\epsilon r_{i}e^{i\pi r/\epsilon }\boldsymbol{\delta }u(r_{i},\phi
^{\prime },\varphi ^{\prime })H_{\epsilon }(r-r_{i}).
\end{eqnarray}%
Here we observed a similar phenomena. Decompose the density variation by
\begin{equation}
\boldsymbol{\delta }u(r,\phi ,\varphi )=\boldsymbol{\delta }A(r,\phi
,\varphi )+\frac{e^{-i\pi r/\epsilon }}{\epsilon }\boldsymbol{\delta }%
B(r,\phi ,\varphi ),
\end{equation}%
and one get the highly-oscillating terms $\boldsymbol{\delta }A\boldsymbol{%
\delta }A$ and $\boldsymbol{\delta }B\boldsymbol{\delta }B$, and the
remaining non-oscillating term $\boldsymbol{\delta }A\boldsymbol{\delta }B$.

The symplectic form is integrated and one can perform a coarse-graining. By
using the coarse-graining function (\ref{eqn 62}), we obtain, in the same
way,$\ $%
\begin{equation}
\boldsymbol{\delta }u_{IR}(x)=\int d^{4}yP_{IR}(x,y)\boldsymbol{\delta }%
u(y)\sim \boldsymbol{\delta }A(x).  \label{eqn 71}
\end{equation}%
This is the emergent new soft mode on the stretched horizon. We have $k\leq
y_{0}^{-1}<k_{\epsilon }=\frac{1}{\epsilon }$. This is because $P_{IR}(k)$
is very small for $k>y_{0}^{-1}$, and $k_{\epsilon }=\frac{1}{\epsilon }$ is
in the range of being removed, hence the highly oscillatory part is removed,
for the coarse-grained data. For observer who has coarse-grained data on the
scale much larger than $\epsilon $, he or she sees the effect of
nonoscillating $\boldsymbol{\delta }A\boldsymbol{\delta }B$. The resulting
phenomena and related conclusions are similar to \cite%
{Balasubramanian:2018yjq}, hence we refer the readers to \cite%
{Balasubramanian:2018yjq} for detailed expositions.

\section{Overview and summary of solution techniques}

\label{sec F}

The solutions come with different techniques. There is an overall solution
space of the smooth geometries. Different solution techniques describe
various regimes of the solution space.

One technique is the solutions of linearized Monge-Ampere equations with
sources. The linearized equations are Laplace-type equations with sources.
The core or source of the linearized solutions can be viewed as the limit of
collapsing of isolated small droplets.

Another technique is dilute distribution and extreme-ratio solutions, where
there are multiple small droplets and can include the central droplet. The
blowing-down of these solutions along the cycles of smooth droplets is
related to the solutions of linearized equations with sources.

Another technique is the wavy deformation of droplets \cite%
{Grant:2005qc,Maoz:2005nk}, which include general cases of wavy deformation
of a general set of droplets. These solutions are also related to volume
preserving deformations. These solutions are used to obtain the symplectic
forms of the configuration space of general solutions, including general
droplets.

Another technique is the stretched droplet space solutions, where coarse
distributions are placed at a stretched surface $y=y_{c}$ away from the $y=0$
surface. The linearized technique can be combined with this technique,
because we can place this stretched surface outside all the enclosure
surfaces of each droplets.


\begin{thebibliography}{999}
\bibitem{Beisert:2010jr} N.~Beisert, C.~Ahn, L.~F.~Alday, Z.~Bajnok,
J.~M.~Drummond, L.~Freyhult, N.~Gromov, R.~A.~Janik, V.~Kazakov and
T.~Klose, \textit{et al.} ``Review of AdS/CFT Integrability: An Overview,''
Lett. Math. Phys. \textbf{99} (2012), 3 
[arXiv:1012.3982 [hep-th]].

\bibitem{Rangamani:2016dms} M.~Rangamani and T.~Takayanagi, ``Holographic
Entanglement Entropy,'' Lect. Notes Phys. \textbf{931} (2017), 1, 2017,
[arXiv:1609.01287 [hep-th]].

\bibitem{Horowitz:2006ct} G.~T.~Horowitz and J.~Polchinski,
\textquotedblleft Gauge/gravity duality,\textquotedblright\ In: D. Oriti,
Approaches to quantum gravity, Cambridge University Press [gr-qc/0602037].

\bibitem{McGreevy:2000cw} J.~McGreevy, L.~Susskind and N.~Toumbas,
``Invasion of the giant gravitons from Anti-de Sitter space,'' JHEP \textbf{%
06} (2000), 008 
[arXiv:hep-th/0003075 [hep-th]].

\bibitem{Grisaru:2000zn} M.~T.~Grisaru, R.~C.~Myers and O.~Tafjord, ``SUSY
and goliath,'' JHEP \textbf{08} (2000), 040
[arXiv:hep-th/0008015 [hep-th]].

\bibitem{Hashimoto:2000zp} A.~Hashimoto, S.~Hirano and N.~Itzhaki, ``Large
branes in AdS and their field theory dual,'' JHEP \textbf{08} (2000), 051
[arXiv:hep-th/0008016 [hep-th]].

\bibitem{Corley:2001zk} S.~Corley, A.~Jevicki and S.~Ramgoolam, ``Exact
correlators of giant gravitons from dual N=4 SYM theory,'' Adv. Theor. Math.
Phys. \textbf{5} (2002), 809-839 
[arXiv:hep-th/0111222 [hep-th]].

\bibitem{Berenstein:2004kk} D.~Berenstein, ``A Toy model for the AdS / CFT
correspondence,'' JHEP \textbf{07} (2004), 018
[arXiv:hep-th/0403110 [hep-th]].

\bibitem{Balasubramanian:2001nh} V.~Balasubramanian, M.~Berkooz, A.~Naqvi
and M.~J.~Strassler, ``Giant gravitons in conformal field theory,'' JHEP
\textbf{04} (2002), 034 
[arXiv:hep-th/0107119 [hep-th]].

\bibitem{Lin:2004nb} H.~Lin, O.~Lunin and J.~M.~Maldacena, ``Bubbling AdS
space and 1/2 BPS geometries,'' JHEP \textbf{10} (2004), 025
[arXiv:hep-th/0409174 [hep-th]].

\bibitem{Gaiotto:2021xce} D.~Gaiotto and J.~H.~Lee, ``The Giant Graviton
Expansion,'' [arXiv:2109.02545 [hep-th]].

\bibitem{Imamura:2022aua} Y.~Imamura, ``Analytic continuation for giant
gravitons,'' PTEP \textbf{2022} (2022) no.10, 103B02
[arXiv:2205.14615 [hep-th]].

\bibitem{Budzik:2021fyh} K.~Budzik and D.~Gaiotto, ``Giant gravitons in
twisted holography,'' [arXiv:2106.14859 [hep-th]].

\bibitem{Murthy:2022ien} S.~Murthy, ``Unitary matrix models, free fermions,
and the giant graviton expansion,''
[arXiv:2202.06897 [hep-th]].

\bibitem{Eniceicu:2023uvd} D.~S.~Eniceicu, ``Comments on the Giant-Graviton
Expansion of the Superconformal Index,'' [arXiv:2302.04887 [hep-th]].

\bibitem{Liu:2022olj} J.~T.~Liu and N.~J.~Rajappa, ``Finite N indices and
the giant graviton expansion,'' 
[arXiv:2212.05408 [hep-th]].

\bibitem{Choi:2022ovw} S.~Choi, S.~Kim, E.~Lee and J.~Lee, ``From giant
gravitons to black holes,'' 
[arXiv:2207.05172 [hep-th]].

\bibitem{Beccaria:2023hip} M.~Beccaria and A.~Cabo-Bizet, ``Large black hole
entropy from the giant brane expansion,'' [arXiv:2308.05191 [hep-th]].

\bibitem{Berenstein:2017abm} D.~Berenstein and A.~Miller, ``Superposition
induced topology changes in quantum gravity,'' JHEP \textbf{11} (2017), 121
[arXiv:1702.03011 [hep-th]].

\bibitem{Balasubramanian:2005mg} V.~Balasubramanian, J.~de Boer, V.~Jejjala
and J.~Simon, ``The Library of Babel: On the origin of gravitational
thermodynamics,'' JHEP \textbf{12} (2005), 006
[arXiv:hep-th/0508023 [hep-th]].

\bibitem{Grant:2005qc} L.~Grant, L.~Maoz, J.~Marsano, K.~Papadodimas and
V.~S.~Rychkov, \textquotedblleft Minisuperspace quantization of Bubbling AdS
and free fermion droplets,\textquotedblright\ JHEP \textbf{08} (2005), 025
[arXiv:hep-th/0505079 [hep-th]].

\bibitem{Maoz:2005nk} L.~Maoz and V.~S.~Rychkov, ``Geometry quantization
from supergravity: The Case of Bubbling AdS,'' JHEP \textbf{08} (2005), 096
[arXiv:hep-th/0508059 [hep-th]].

\bibitem{Bhattacharyya:2008rb} R.~Bhattacharyya, S.~Collins and R.~de Mello
Koch, ``Exact Multi-Matrix Correlators,'' JHEP \textbf{03} (2008), 044
[arXiv:0801.2061 [hep-th]].

\bibitem{Bhattacharyya:2008xy} R.~Bhattacharyya, R.~de Mello Koch and
M.~Stephanou, ``Exact Multi-Restricted Schur Polynomial Correlators,'' JHEP
\textbf{06} (2008), 101 
[arXiv:0805.3025 [hep-th]].

\bibitem{deMelloKoch:2012sie} R.~de Mello Koch, P.~Diaz and N.~Nokwara,
``Restricted Schur Polynomials for Fermions and integrability in the su(2$|$%
3) sector,'' JHEP \textbf{03} (2013), 173 
[arXiv:1212.5935 [hep-th]].

\bibitem{Brown:2007xh} T.~W.~Brown, P.~J.~Heslop and S.~Ramgoolam,
``Diagonal multi-matrix correlators and BPS operators in N=4 SYM,'' JHEP
\textbf{02} (2008), 030 
[arXiv:0711.0176 [hep-th]].

\bibitem{Berenstein:2022srd} D.~Berenstein and S.~Wang, ``BPS coherent
states and localization,'' JHEP \textbf{08} (2022), 164
[arXiv:2203.15820 [hep-th]].

\bibitem{Berenstein:2005aa} D.~Berenstein, ``Large N BPS states and emergent
quantum gravity,'' JHEP \textbf{01} (2006), 125
[arXiv:hep-th/0507203 [hep-th]].

\bibitem{Dhar:2005su} A.~Dhar, G.~Mandal and M.~Smedback, ``From gravitons
to giants,'' JHEP \textbf{03} (2006), 031
[arXiv:hep-th/0512312 [hep-th]].

\bibitem{Kimura:2007wy} Y.~Kimura and S.~Ramgoolam, ``Branes, anti-branes
and brauer algebras in gauge-gravity duality,'' JHEP \textbf{11} (2007), 078
[arXiv:0709.2158 [hep-th]].

\bibitem{deMelloKoch:2007nbd} R.~de Mello Koch, J.~Smolic and M.~Smolic,
``Giant Gravitons - with Strings Attached (II),'' JHEP \textbf{09} (2007),
049 
[arXiv:hep-th/0701067 [hep-th]].

\bibitem{Caputa:2013hr} P.~Caputa, R.~de Mello Koch and P.~Diaz, ``A basis
for large operators in N=4 SYM with orthogonal gauge group,'' JHEP \textbf{03%
} (2013), 041 
[arXiv:1301.1560 [hep-th]].

\bibitem{Suzuki:2020oce} R.~Suzuki, ``Three-point functions in $\mathcal{N} $
= 4 SYM at finite N$_{c}$ and background independence,'' JHEP \textbf{05}
(2020), 118 
[arXiv:2002.07216 [hep-th]].

\bibitem{Pasukonis:2013ts} J.~Pasukonis and S.~Ramgoolam, ``Quivers as
Calculators: Counting, Correlators and Riemann Surfaces,'' JHEP \textbf{04}
(2013), 094 
[arXiv:1301.1980 [hep-th]].

\bibitem{Lewis-Brown:2020nmg} C.~Lewis-Brown and S.~Ramgoolam, ``Quarter-BPS
states, multi-symmetric functions and set partitions,'' JHEP \textbf{03}
(2021), 153 
[arXiv:2007.01734 [hep-th]].

\bibitem{Bissi:2011dc} A.~Bissi, C.~Kristjansen, D.~Young and K.~Zoubos,
``Holographic three-point functions of giant gravitons,'' JHEP \textbf{06}
(2011), 085 
[arXiv:1103.4079 [hep-th]].

\bibitem{Bak:2011yy} D.~Bak, B.~Chen and J.~B.~Wu, ``Holographic Correlation
Functions for Open Strings and Branes,'' JHEP \textbf{06} (2011), 014
[arXiv:1103.2024 [hep-th]].

\bibitem{Caputa:2012yj} P.~Caputa, R.~de Mello Koch and K.~Zoubos,
``Extremal versus Non-Extremal Correlators with Giant Gravitons,'' JHEP
\textbf{08} (2012), 143 
[arXiv:1204.4172 [hep-th]].

\bibitem{Lin:2012ey} H.~Lin, ``Giant gravitons and correlators,'' JHEP
\textbf{12} (2012), 011 
[arXiv:1209.6624 [hep-th]].

\bibitem{Jiang:2019xdz} Y.~Jiang, S.~Komatsu and E.~Vescovi,
\textquotedblleft Structure constants in $\mathcal{N}$ = 4 SYM at finite
coupling as worldsheet g-function,\textquotedblright\ JHEP \textbf{07}
(2020), 037 
[arXiv:1906.07733 [hep-th]].

\bibitem{Chen:2019gsb} G.~Chen, R.~de Mello Koch, M.~Kim and H.~J.~R.~Van
Zyl, ``Absorption of closed strings by giant gravitons,'' JHEP \textbf{10}
(2019), 133 
[arXiv:1908.03553 [hep-th]].

\bibitem{Yang:2021kot} P.~Yang, Y.~Jiang, S.~Komatsu and J.~B.~Wu,
``D-branes and orbit average,'' SciPost Phys. \textbf{12} (2022) no.2, 055
[arXiv:2103.16580 [hep-th]].

\bibitem{Holguin:2022zii} A.~Holguin and W.~W.~Weng, ``Orbit averaging
coherent states: holographic three-point functions of AdS giant gravitons,''
JHEP \textbf{05} (2023), 167 
[arXiv:2211.03805 [hep-th]].

\bibitem{Abajian:2023jye} J.~Abajian, F.~Aprile, R.~C.~Myers and P.~Vieira,
``Holography and correlation functions of huge operators: spacetime
bananas,'' [arXiv:2306.15105 [hep-th]].

\bibitem{Holguin:2023orq} A.~Holguin, ``1/2 BPS structure constants and
random matrices,'' JHEP \textbf{12} (2023), 046
[arXiv:2305.06390 [hep-th]].

\bibitem{Chen:2007du} B.~Chen, S.~Cremonini, A.~Donos, F.~L.~Lin, H.~Lin,
J.~T.~Liu, D.~Vaman and W.~Y.~Wen, ``Bubbling AdS and droplet descriptions
of BPS geometries in IIB supergravity,'' JHEP \textbf{10} (2007), 003
[arXiv:0704.2233 [hep-th]].

\bibitem{Lunin:2008tf} O.~Lunin, ``Brane webs and 1/4-BPS geometries,'' JHEP
\textbf{09} (2008), 028 
[arXiv:0802.0735 [hep-th]].

\bibitem{Lin:2010nd} H.~Lin, ``Studies on 1/4 BPS and 1/8 BPS geometries,''
[arXiv:1008.5307 [hep-th]].

\bibitem{Kimura:2011df} Y.~Kimura and H.~Lin, ``Young diagrams, Brauer
algebras, and bubbling geometries,'' JHEP \textbf{01} (2012), 121
[arXiv:1109.2585 [hep-th]].

\bibitem{Gauntlett:2006ns} J.~P.~Gauntlett, N.~Kim and D.~Waldram,
``Supersymmetric AdS(3), AdS(2) and Bubble Solutions,'' JHEP \textbf{04}
(2007), 005 
[arXiv:hep-th/0612253 [hep-th]].

\bibitem{Kim:2005ez} N.~Kim, ``AdS(3) solutions of IIB supergravity from
D3-branes,'' JHEP \textbf{01} (2006), 094
[arXiv:hep-th/0511029 [hep-th]].

\bibitem{Gava:2006pu} E.~Gava, G.~Milanesi, K.~S.~Narain and M.~O'Loughlin,
``1/8 BPS states in AdS/CFT,'' JHEP \textbf{05} (2007), 030
[arXiv:hep-th/0611065 [hep-th]].

\bibitem{Liu:2007xj} J.~T.~Liu, H.~Lu, C.~N.~Pope and J.~F.~Vazquez-Poritz,
``Bubbling AdS black holes,'' JHEP \textbf{10} (2007), 030
[arXiv:hep-th/0703184 [hep-th]].

\bibitem{Chong:2004ce} Z.~W.~Chong, H.~Lu and C.~N.~Pope, ``BPS geometries
and AdS bubbles,'' Phys. Lett. B \textbf{614} (2005), 96-103
[arXiv:hep-th/0412221 [hep-th]].

\bibitem{Lunin:2007ab} O.~Lunin, ``1/2-BPS states in M theory and defects in
the dual CFTs,'' JHEP \textbf{10} (2007), 014
[arXiv:0704.3442 [hep-th]].

\bibitem{Liu:2004ru} J.~T.~Liu, D.~Vaman and W.~Y.~Wen, ``Bubbling 1/4 BPS
solutions in type IIB and supergravity reductions on S*n x S*n,'' Nucl.
Phys. B \textbf{739} (2006), 285-310 
[arXiv:hep-th/0412043 [hep-th]].

\bibitem{Balasubramanian:2018yjq} V.~Balasubramanian, D.~Berenstein,
A.~Lewkowycz, A.~Miller, O.~Parrikar and C.~Rabideau, ``Emergent classical
spacetime from microstates of an incipient black hole,'' JHEP \textbf{01}
(2019), 197 
[arXiv:1810.13440 [hep-th]].

\bibitem{Mukhi:2005cv} S.~Mukhi and M.~Smedback, ``Bubbling orientifolds,''
JHEP \textbf{08} (2005), 005 
[arXiv:hep-th/0506059 [hep-th]].

\bibitem{Fiol:2014fla} B.~Fiol, B.~Garolera and G.~Torrents, ``Exact probes
of orientifolds,'' JHEP \textbf{09} (2014), 169
[arXiv:1406.5129 [hep-th]].

\bibitem{Lewis-Brown:2018dje} C.~Lewis-Brown and S.~Ramgoolam, ``BPS
operators in $\mathcal{N}=4$ $SO(N)$ super Yang-Mills theory: plethysms,
dominoes and words,'' JHEP \textbf{11} (2018), 035
[arXiv:1804.11090 [hep-th]].

\bibitem{Holguin:2022drf} A.~Holguin and S.~Wang, ``Giant gravitons,
Harish-Chandra integrals, and BPS states in symplectic and orthogonal $%
\mathcal{N} $ = 4 SYM,'' JHEP \textbf{10} (2022), 078
[arXiv:2206.00020 [hep-th]].

\bibitem{Giombi:2020kvo} S.~Giombi and B.~Offertaler, ``Wilson loops in $%
\mathcal{N} $ = 4 SO(N) SYM and D-branes in AdS$_{5}$ \texttimes{} $\mathbb{R%
}$$\mathbb{P}$$^{5}$,'' JHEP \textbf{10} (2021), 016
[arXiv:2006.10852 [hep-th]].

\bibitem{Caputa:2013vla} P.~Caputa, R.~de Mello Koch and P.~Diaz,
``Operators, Correlators and Free Fermions for SO(N) and Sp(N),'' JHEP
\textbf{06} (2013), 018 
[arXiv:1303.7252 [hep-th]].

\bibitem{Bhattacharyya:2010yg} S.~Bhattacharyya, S.~Minwalla and
K.~Papadodimas, ``Small Hairy Black Holes in $AdS_5$x$S^5$,'' JHEP \textbf{11%
} (2011), 035 
[arXiv:1005.1287 [hep-th]].

\bibitem{Kunduri:2007qy} H.~K.~Kunduri and J.~Lucietti, ``Near-horizon
geometries of supersymmetric AdS(5) black holes,'' JHEP \textbf{12} (2007),
015 
[arXiv:0708.3695 [hep-th]].

\bibitem{Skenderis:2007yb} K.~Skenderis and M.~Taylor, ``Anatomy of bubbling
solutions,'' JHEP \textbf{09} (2007), 019
[arXiv:0706.0216 [hep-th]].

\bibitem{Christodoulou:2016nej} A.~Christodoulou and K.~Skenderis,
``Holographic Construction of Excited CFT States,'' JHEP \textbf{04} (2016),
096 
[arXiv:1602.02039 [hep-th]].

\bibitem{Armas:2013ota} J.~Armas, N.~A.~Obers and A.~V.~Pedersen,
``Null-Wave Giant Gravitons from Thermal Spinning Brane Probes,'' JHEP
\textbf{10} (2013), 109 
[arXiv:1306.2633 [hep-th]].

\bibitem{Kemp:2019log} G.~Kemp and S.~Ramgoolam, ``BPS states, conserved
charges and centres of symmetric group algebras,'' JHEP \textbf{01} (2020),
[arXiv:1911.11649 [hep-th]].

\bibitem{Geloun:2023zqa} J.~B.~Geloun and S.~Ramgoolam, ``The quantum
detection of projectors in finite-dimensional algebras and holography,''
JHEP \textbf{05} (2023), 191 
[arXiv:2303.12154 [quant-ph]].

\bibitem{Balasubramanian:2006jt} V.~Balasubramanian, B.~Czech, K.~Larjo and
J.~Simon, ``Integrability versus information loss: A Simple example,'' JHEP
\textbf{11} (2006), 001 
[arXiv:hep-th/0602263 [hep-th]].

\bibitem{Balasubramanian:2007zt} V.~Balasubramanian, B.~Czech, K.~Larjo,
D.~Marolf and J.~Simon, ``Quantum geometry and gravitational entropy,'' JHEP
\textbf{12} (2007), 067 
[arXiv:0705.4431 [hep-th]].

\bibitem{Balasubramanian:2007qv} V.~Balasubramanian, B.~Czech, V.~E.~Hubeny,
K.~Larjo, M.~Rangamani and J.~Simon, ``Typicality versus thermality: An
Analytic distinction,'' Gen. Rel. Grav. \textbf{40} (2008), 1863-1890
[arXiv:hep-th/0701122 [hep-th]].

\bibitem{Biswas:2006tj} I.~Biswas, D.~Gaiotto, S.~Lahiri and S.~Minwalla,
``Supersymmetric states of N=4 Yang-Mills from giant gravitons,'' JHEP
\textbf{12} (2007), 006 
[arXiv:hep-th/0606087 [hep-th]].

\bibitem{Mandal:2006tk} G.~Mandal and N.~V.~Suryanarayana, ``Counting
1/8-BPS dual-giants,'' JHEP \textbf{03} (2007), 031
[arXiv:hep-th/0606088 [hep-th]].

\bibitem{Kim:2006he} S.~Kim and K.~M.~Lee, ``1/16-BPS Black Holes and Giant
Gravitons in the AdS(5) x S*5 Space,'' JHEP \textbf{12} (2006), 077
[arXiv:hep-th/0607085 [hep-th]].

\bibitem{Paul:2023rka} H.~Paul, E.~Perlmutter and H.~Raj, ``Exact large
charge in $\mathcal{N} $ = 4 SYM and semiclassical string theory,'' JHEP
\textbf{08} (2023), 078 
[arXiv:2303.13207 [hep-th]].

\bibitem{Caetano:2023zwe} J.~Caetano, S.~Komatsu and Y.~Wang, ``Large Charge
't Hooft Limit of $\mathcal{N}=4$ Super-Yang-Mills,'' [arXiv:2306.00929
[hep-th]].

\bibitem{Li:2019tpf} Y.~Z.~Li, Z.~F.~Mai, H.~Lu, ``Holographic OPE
Coefficients from AdS Black Holes with Matters,'' JHEP \textbf{09} (2019),
001 
[arXiv:1905.09302 [hep-th]].

\bibitem{deMelloKoch:2009jc} R.~de Mello Koch, T.~K.~Dey, N.~Ives and
M.~Stephanou, ``Correlators Of Operators with a Large R-charge,'' JHEP
\textbf{08} (2009), 083 
[arXiv:0905.2273 [hep-th]].

\bibitem{Berenstein:2020jen} D.~Berenstein and A.~Holguin, ``Open giant
magnons on LLM geometries,'' JHEP \textbf{01} (2021), 080
[arXiv:2010.02236 [hep-th]].

\bibitem{Chen:2007gh} H.~Y.~Chen, D.~H.~Correa and G.~A.~Silva, ``Geometry
and topology of bubble solutions from gauge theory,'' Phys. Rev. D \textbf{76%
} (2007), 026003 
[arXiv:hep-th/0703068 [hep-th]].

\bibitem{deMelloKoch:2018ert} R.~de Mello Koch, J.~H.~Huang and
L.~Tribelhorn, ``Exciting LLM Geometries,'' JHEP \textbf{07} (2018), 146
[arXiv:1806.06586 [hep-th]].

\bibitem{Yamaguchi:2006te} S.~Yamaguchi, ``Bubbling geometries for half BPS
Wilson lines,'' Int. J. Mod. Phys. A \textbf{22} (2007), 1353-1374
[arXiv:hep-th/0601089 [hep-th]].

\bibitem{Lunin:2006xr} O.~Lunin, ``On gravitational description of Wilson
lines,'' JHEP \textbf{0606} (2006) 026 
[hep-th/0604133]. 

\bibitem{Gomis:2006sb} J.~Gomis and F.~Passerini, ``Holographic Wilson
Loops,'' JHEP \textbf{08} (2006), 074 
[arXiv:hep-th/0604007 [hep-th]].

\bibitem{DHoker:2007mci} E.~D'Hoker, J.~Estes and M.~Gutperle, ``Gravity
duals of half-BPS Wilson loops,'' JHEP \textbf{0706} (2007) 063
[arXiv:0705.1004 [hep-th]].

\bibitem{Aguilera-Damia:2017znn} J.~Aguilera-Damia, D.~H.~Correa, F.~Fucito,
V.~I.~Giraldo-Rivera, J.~F.~Morales and L.~A.~Pando Zayas, ``Strings in
Bubbling Geometries and Dual Wilson Loop Correlators,'' JHEP \textbf{12}
(2017), 109 
[arXiv:1709.03569 [hep-th]].

\bibitem{Yamaguchi:2006tq} S.~Yamaguchi, ``Wilson loops of anti-symmetric
representation and D5-branes,'' JHEP \textbf{05} (2006), 037
[arXiv:hep-th/0603208 [hep-th]].

\bibitem{Gentle:2015jma} S.~A.~Gentle, M.~Gutperle and C.~Marasinou,
``Entanglement entropy of Wilson surfaces from bubbling geometries in
M-theory,'' JHEP \textbf{08} (2015), 019 
[arXiv:1506.00052 [hep-th]].

\bibitem{Okuda:2007kh} T.~Okuda, ``A Prediction for bubbling geometries,''
JHEP \textbf{01} (2008), 003 
[arXiv:0708.3393 [hep-th]].

\bibitem{Giombi:2009ds} S.~Giombi and V.~Pestun, ``Correlators of local
operators and 1/8 BPS Wilson loops on S*2 from 2d YM and matrix models,''
JHEP \textbf{10} (2010), 033 
[arXiv:0906.1572 [hep-th]].

\bibitem{Hartnoll:2006is} S.~A.~Hartnoll and S.~P.~Kumar, ``Higher rank
Wilson loops from a matrix model,'' JHEP \textbf{08} (2006), 026
[arXiv:hep-th/0605027 [hep-th]].

\bibitem{Gomis:2008qa} J.~Gomis, S.~Matsuura, T.~Okuda and D.~Trancanelli,
``Wilson loop correlators at strong coupling: From matrices to bubbling
geometries,'' JHEP \textbf{08} (2008), 068
[arXiv:0807.3330 [hep-th]].

\bibitem{Gukov:2008sn} S.~Gukov and E.~Witten, ``Rigid Surface Operators,''
Adv. Theor. Math. Phys. \textbf{14} (2010) no.1, 87-178
[arXiv:0804.1561 [hep-th]].

\bibitem{Gomis:2007fi} J.~Gomis and S.~Matsuura, ``Bubbling surface
operators and S-duality,'' JHEP \textbf{06} (2007), 025
[arXiv:0704.1657 [hep-th]].

\bibitem{Gentle:2015ruo} S.~A.~Gentle, M.~Gutperle and C.~Marasinou,
``Holographic entanglement entropy of surface defects,'' JHEP \textbf{04}
(2016), 067 
[arXiv:1512.04953 [hep-th]].

\bibitem{Lin:2005nh} H.~Lin and J.~M.~Maldacena, ``Fivebranes from gauge
theory,'' Phys. Rev. D \textbf{74} (2006), 084014
[arXiv:hep-th/0509235 [hep-th]].

\bibitem{Drukker:2008wr} N.~Drukker, J.~Gomis and S.~Matsuura, ``Probing N=4
SYM With Surface Operators,'' JHEP \textbf{10} (2008), 048
[arXiv:0805.4199 [hep-th]].

\bibitem{Asano:2014eca} Y.~Asano, G.~Ishiki and S.~Shimasaki, ``Emergent
bubbling geometries in gauge theories with SU(2$|$4) symmetry,'' JHEP
\textbf{09} (2014), 137 
[arXiv:1406.1337 [hep-th]].

\bibitem{Asano:2012zt} Y.~Asano, G.~Ishiki, T.~Okada and S.~Shimasaki,
``Exact results for perturbative partition functions of theories with SU(2$|$%
4) symmetry,'' JHEP \textbf{02} (2013), 148 
[arXiv:1211.0364 [hep-th]].

\bibitem{Jejjala:2008jy} V.~Jejjala, D.~Minic, Y.~J.~Ng and C.~H.~Tze,
``Turbulence and Holography,'' Class. Quant. Grav. \textbf{25} (2008),
225012 
[arXiv:0806.0030 [hep-th]].

\bibitem{Caldarelli:2004mz} M.~M.~Caldarelli, D.~Klemm and P.~J.~Silva,
``Chronology protection in anti-de Sitter,'' Class. Quant. Grav. \textbf{22}
(2005), 3461-3466 
[arXiv:hep-th/0411203 [hep-th]].

\bibitem{Berenstein:2007wz} D.~Berenstein and R.~Cotta, ``A Monte-Carlo
study of the AdS/CFT correspondence: An Exploration of quantum gravity
effects,'' JHEP \textbf{04} (2007), 071 
[arXiv:hep-th/0702090 [hep-th]].

\bibitem{Simon:2018laf} J.~Simon, ``Correlations vs connectivity in
R-charge,'' JHEP \textbf{10} (2018), 048 
[arXiv:1805.11279 [hep-th]].

\bibitem{Berenstein:2023srv} D.~Berenstein and K.~Yan, ``The endpoint of
partial deconfinement,'' [arXiv:2307.06122 [hep-th]].

\bibitem{Ahmadain:2022gfw} A.~Ahmadain, A.~Frenkel, K.~Ray and R.~M.~Soni,
``Boundary Description of Microstates of the Two-Dimensional Black Hole,''
[arXiv:2210.11493 [hep-th]].

\bibitem{Balasubramanian:2007bs} V.~Balasubramanian, J.~de Boer, V.~Jejjala
and J.~Simon, ``Entropy of near-extremal black holes in AdS(5),'' JHEP
\textbf{05} (2008), 067 
[arXiv:0707.3601 [hep-th]].

\bibitem{Fareghbal:2008ar} R.~Fareghbal, C.~N.~Gowdigere, A.~E.~Mosaffa and
M.~M.~Sheikh-Jabbari, ``Nearing Extremal Intersecting Giants and New
Decoupled Sectors in N = 4 SYM,'' JHEP \textbf{08} (2008), 070
[arXiv:0801.4457 [hep-th]].

\bibitem{Myers:2001aq} R. C. Myers and O. Tafjord, ``Superstars and giant
gravitons,'' JHEP \textbf{11} (2001), 009
[arXiv:hep-th/0109127 [hep-th]].

\bibitem{Cordova:2016emh} C.~Cordova, T.~T.~Dumitrescu and K.~Intriligator,
``Multiplets of Superconformal Symmetry in Diverse Dimensions,'' JHEP
\textbf{03} (2019), 163 
[arXiv:1612.00809 [hep-th]].

\bibitem{Caetano:2020ofu} J.~Caetano, W.~Peelaers and L.~Rastelli,
``Maximally supersymmetric RG flows in 4D and integrability,'' JHEP \textbf{%
12} (2021), 119 
[arXiv:2006.04792 [hep-th]].

\bibitem{Cordova:2016xhm} C.~Cordova, T.~T.~Dumitrescu and K.~Intriligator,
``Deformations of Superconformal Theories,'' JHEP \textbf{11} (2016), 135
[arXiv:1602.01217 [hep-th]].

\bibitem{Roychowdhury:2023hvq} D.~Roychowdhury, ``Matrix model correlators
from non-Abelian T-dual of $AdS_5 \times S^5 $,'' [arXiv:2310.10210
[hep-th]].

\bibitem{Nomura:2019qps} Y.~Nomura, ``Spacetime and Universal Soft Modes ---
Black Holes and Beyond,'' Phys. Rev. D \textbf{101} (2020), 066024
[arXiv:1908.05728 [hep-th]].

\bibitem{Witten:2012bh} E.~Witten, ``Superstring Perturbation Theory
Revisited,'' [arXiv:1209.5461 [hep-th]].

\bibitem{Berenstein:2023vtd} D.~Berenstein, E.~Maderazo, R.~Mancilla and
A.~Ramirez, ``Chaotic LLM billiards,'' [arXiv:2305.19321 [hep-th]].

\bibitem{Mayerson:2020tpn} D.~R.~Mayerson, ``Fuzzballs and Observations,''
Gen. Rel. Grav. \textbf{52} (2020) no.12, 115
[arXiv:2010.09736 [hep-th]].

\bibitem{Mathur:2009hf} S.~D.~Mathur,
Class. Quant. Grav. \textbf{26} (2009), 224001
[arXiv:0909.1038 [hep-th]].

\bibitem{Bianchi:2017sds} M.~Bianchi, D.~Consoli and J.~F.~Morales,
``Probing Fuzzballs with Particles, Waves and Strings,'' JHEP \textbf{06}
(2018), 157 
[arXiv:1711.10287 [hep-th]].

\bibitem{Chakrabarty:2021sff} B.~Chakrabarty, S.~Rawash and D.~Turton,
``Shockwaves in black hole microstate geometries,'' JHEP \textbf{02} (2022),
202 
[arXiv:2112.08378 [hep-th]].

\bibitem{Heidmann:2023ojf} P.~Heidmann, N.~Speeney, E.~Berti and I.~Bah,
``Cavity effect in the quasinormal mode spectrum of topological stars,''
Phys. Rev. D \textbf{108} (2023) 024021 
[arXiv:2305.14412 [gr-qc]].

\bibitem{Bacchini:2021fig} F.~Bacchini, D.~R.~Mayerson, B.~Ripperda,
J.~Davelaar, H.~Olivares, T.~Hertog and B.~Vercnocke, ``Fuzzball Shadows:
Emergent Horizons from Microstructure,'' Phys. Rev. Lett. \textbf{127}
(2021), 171601 
[arXiv:2103.12075 [hep-th]].

\bibitem{Bena:2022ldq} I.~Bena, E.~J.~Martinec, S.~D.~Mathur and
N.~P.~Warner, ``Snowmass White Paper: Micro- and Macro-Structure of Black
Holes,'' [arXiv:2203.04981 [hep-th]].

\bibitem{deMelloKoch:2020jmf} R.~de Mello Koch, E.~Gandote and A.~L.~Mahu,
``Scrambling in Yang-Mills,'' JHEP \textbf{01} (2021), 058
[arXiv:2008.12409 [hep-th]].

\bibitem{McLoughlin:2020zew} T.~McLoughlin, R.~Pereira and A.~Spiering,
``Quantum chaos in perturbative super-Yang-Mills Theory,'' SciPost Phys.
\textbf{14} (2023) no.3, 049 
[arXiv:2011.04633 [hep-th]].

\bibitem{Behrndt:1998jd} K.~Behrndt, M.~Cvetic and W.~A.~Sabra, ``Nonextreme
black holes of five-dimensional N=2 AdS supergravity,'' Nucl. Phys. B
\textbf{553} (1999), 317-332 
[arXiv:hep-th/9810227 [hep-th]].

\bibitem{Behrndt:1998ns} K.~Behrndt, A.~H.~Chamseddine and W.~A.~Sabra,
``BPS black holes in N=2 five-dimensional AdS supergravity,'' Phys. Lett. B
\textbf{442} (1998), 97-101 
[arXiv:hep-th/9807187 [hep-th]].

\bibitem{Berenstein:2022cju} D.~Berenstein and A.~Holguin, ``String
junctions suspended between giants,'' JHEP \textbf{11} (2022), 085
[arXiv:2202.11729 [hep-th]].

\bibitem{Holguin:2021qes} A.~Holguin, ``Giant Gravitons Intersecting at
Angles from Integrable Spin Chains,'' [arXiv:2111.05981 [hep-th]].

\bibitem{Baiguera:2021hky} S.~Baiguera, T.~Harmark and Y.~Lei, ``Spin Matrix
Theory in near $\frac{1}{8} $-BPS corners of $\mathcal{N} $ = 4
super-Yang-Mills,'' JHEP \textbf{02} (2022), 191
[arXiv:2111.10149 [hep-th]].

\bibitem{Baiguera:2022pll} S.~Baiguera, T.~Harmark and Y.~Lei, ``The
Panorama of Spin Matrix theory,'' JHEP \textbf{04} (2023), 075
[arXiv:2211.16519 [hep-th]].

\bibitem{Baiguera:2020mgk} S.~Baiguera, T.~Harmark, Y.~Lei and
N.~Wintergerst, ``Symmetry structure of the interactions in near-BPS corners
of $\mathcal{N} = 4$ super-Yang-Mills,'' JHEP \textbf{04} (2021), 029
[arXiv:2012.08532 [hep-th]].

\bibitem{Harmark:2016cjq} T.~Harmark, ``Interacting Giant Gravitons from
Spin Matrix Theory,'' Phys. Rev. D \textbf{94} (2016) no.6, 066001
[arXiv:1606.06296 [hep-th]].

\bibitem{Lin:2022wdr} H.~Lin, ``Coherent state excitations and string-added
coherent states in gauge-gravity correspondence,'' Nucl. Phys. B \textbf{986}
(2023), 116066 
[arXiv:2206.06524 [hep-th]].

\bibitem{deMelloKoch:2012ck} R.~de Mello Koch and S.~Ramgoolam, ``A double
coset ansatz for integrability in AdS/CFT,'' JHEP \textbf{06} (2012), 083
[arXiv:1204.2153 [hep-th]]

\bibitem{Carlson:2011hy} W.~Carlson, R.~de Mello Koch and H.~Lin,
``Nonplanar Integrability,'' JHEP \textbf{03} (2011), 105
[arXiv:1101.5404 [hep-th]].

\bibitem{deMelloKoch:2018tlb} R.~de Mello Koch, M.~Kim and H.~J.~R.~Van~Zyl,
``Integrable Subsectors from Holography,'' JHEP \textbf{05} (2018), 198
[arXiv:1802.01367 [hep-th]].

\bibitem{Berenstein:2008eg} D.~E.~Berenstein, M.~Hanada and S.~A.~Hartnoll,
``Multi-matrix models and emergent geometry,'' JHEP \textbf{02} (2009), 010
[arXiv:0805.4658 [hep-th]].

\bibitem{Kiritsis:2009hu} E.~Kiritsis, ``Dissecting the string theory dual
of QCD,'' Fortsch. Phys. \textbf{57} (2009), 396-417
[arXiv:0901.1772 [hep-th]].

\bibitem{Donos:2005vm} A.~Donos, A.~Jevicki and J.~P.~Rodrigues, ``Matrix
model maps in AdS/CFT,'' Phys. Rev. D \textbf{72} (2005), 125009
[arXiv:hep-th/0507124 [hep-th]].

\bibitem{Correa:2010zj} D.~H.~Correa and M.~Wolf, ``Shaping up BPS States
with Matrix Model Saddle Points,'' J. Phys. A \textbf{43} (2010), 465402
[arXiv:1007.5284 [hep-th]].

\bibitem{Gursoy:2007np} U.~Gursoy, S.~A.~Hartnoll, T.~J.~Hollowood and
S.~P.~Kumar, ``Topology change and new phases in thermal N=4 SYM theory,''
JHEP \textbf{11} (2007), 020 
[arXiv:hep-th/0703100 [hep-th]].

\bibitem{Dutta:2007ws} S.~Dutta and R.~Gopakumar, ``Free fermions and
thermal AdS/CFT,'' JHEP \textbf{03} (2008), 011
[arXiv:0711.0133 [hep-th]].

\bibitem{Berenstein:2006yy} D.~Berenstein and R.~Cotta, ``Aspects of
emergent geometry in the AdS/CFT context,'' Phys. Rev. D \textbf{74} (2006),
026006 
[arXiv:hep-th/0605220 [hep-th]].

\bibitem{Ishii:2008tm} T.~Ishii, G.~Ishiki, S.~Shimasaki and A.~Tsuchiya,
``Fiber Bundles and Matrix Models,'' Phys. Rev. D \textbf{77} (2008), 126015
[arXiv:0802.2782 [hep-th]].

\bibitem{Mandal:2013id} G.~Mandal and T.~Morita, ``Quantum quench in matrix
models: Dynamical phase transitions, Selective equilibration and the
Generalized Gibbs Ensemble,'' JHEP \textbf{10} (2013), 197
[arXiv:1302.0859 [hep-th]].

\bibitem{deMelloKoch:2016whh} R.~de Mello Koch, D.~Gossman, L.~Nkumane and
L.~Tribelhorn, ``Eigenvalue Dynamics for Multimatrix Models,'' Phys. Rev. D
\textbf{96} (2017) no.2, 026011 
[arXiv:1608.00399 [hep-th]].

\bibitem{Holguin:2023naq} A.~Holguin and S.~Wang, ``Multi-matrix correlators
and localization,'' [arXiv:2307.03235 [hep-th]].

\bibitem{Lin:2022gbu} H.~Lin, ``Coherent state operators, giant gravitons,
and gauge-gravity correspondence,'' Annals Phys. \textbf{451} (2023), 169248
[arXiv:2212.14002 [hep-th]].

\bibitem{Berenstein:2008jn} D.~Berenstein, R.~Cotta and R.~Leonardi,
``Numerical tests of AdS/CFT at strong coupling,'' Phys. Rev. D \textbf{78}
(2008), 025008 
[arXiv:0801.2739 [hep-th]].

\bibitem{Puta:} M.~Puta, ``Hamiltonian Mechanical Systems and Geometric
Quantization,'' Mathematics and its applications, v. 260, 1994.

\bibitem{Belin:2018fxe} A.~Belin, A.~Lewkowycz and G.~Sarosi, ``The boundary
dual of the bulk symplectic form,'' Phys. Lett. B \textbf{789} (2019), 71-75
[arXiv:1806.10144 [hep-th]].

\end{thebibliography}
\end{document}